\documentclass[12pt] {article}
\pdfoutput=1
\usepackage[hyphens,spaces,obeyspaces]{url}
\usepackage{amsmath}
\usepackage{datetime}
\usepackage{algorithm}
\usepackage{graphicx}
\usepackage{acronym}
\usepackage{cite}
\usepackage{mystyle}
\usepackage[english]{babel}
\usepackage{caption,subcaption}
\usepackage{tabularx}
\usepackage{enumerate}
\usepackage{marvosym}
\usepackage{makecell}
\usepackage{hyperref}

\AtBeginEnvironment{thm}{\footnotesize}
\AtBeginEnvironment{defn}{\footnotesize}
\providecommand{\keywords}[1]{\textbf{\textit{Keywords---}} #1}

\newsavebox\mybox

{\begin{list}{}%
         {\setlength{\leftmargin}{#1}}%
         \item[]%
}
{\end{list}}

\newdateformat{monthyeardate}{%
  \monthname[\THEMONTH], \THEYEAR}

\begin{document}
\sloppy

\title{\Large{\textbf{Formal FT-based Cause-Consequence Reliability Analysis using Theorem Proving} }}
\author{Mohamed Abdelghany and Sofi\`ene Tahar\vspace*{2em}\\
Department of Electrical and Computer Engineering,\\
Concordia University, Montr\'eal, QC, Canada 
\vspace*{1em}\\
\{m\_eldes,tahar\}@ece.concordia.ca
 \vspace*{3em}\\
\textbf{TECHNICAL REPORT}\\
\date{January 2021}
}
\maketitle

\newpage
\begin{abstract}

Cause-consequence Diagram (CCD) is widely used as a deductive safety analysis technique for decision-making at the critical-system design stage. This approach models the causes of subsystem failures in a highly-critical system and their potential consequences using Fault Tree (FT) and Event Tree~(ET) methods, which are well-known dependability modeling techniques. Paper-and-pencil-based approaches and simulation tools, such as the Monte-Carlo approach, are commonly used to carry out CCD analysis, but lack the ability to rigorously verify essential system reliability properties. In this work, we propose to use formal techniques based on theorem proving for the formal modeling and step-analysis of CCDs  to overcome the inaccuracies of the simulation-based analysis and the error-proneness of informal reasoning by mathematical proofs. In particular, we use the HOL4 theorem prover, which is a computer-based mathematical reasoning tool. To this end, we developed a formalization of CCDs in Higher-Order Logic (HOL), based on the algebraic approach, using HOL4. We demonstrate the practical effectiveness of the proposed CCD formalization by performing the formal reliability analysis of the IEEE 39-bus electrical power network. Also, we formally determine the Forced Outage Rate ($\mathcal{FOR}$) of the power generation units and the network reliability index, i.e., System Average Interruption Duration Index~($\mathcal{SAIDI}$). To assess the accuracy of our proposed approach, we compare our results with those obtained with MATLAB Monte-Carlo Simulation (MCS) as well as other state-of-the-art approaches for subsystem-level reliability~analysis. \\ 
\end{abstract}

\pagebreak
\section{Introduction}
\label{sec:introduction}
Nowadays, in many safety-critical systems, which are prevalent, e.g. in smart grids \cite{fang2011smart} and automotive industry \cite{rahman2004power}, a catastrophic accident may happen due to coincidence of sudden events and/or failures of specific subsystem components. These undesirable accidents may result in loss of profits and sometimes severe fatalities. Therefore, the central inquiry, in many critical-systems, where safety is of the utmost importance, is to identify the possible consequences given that one or more components could fail at a subsystem level on the entire system. For that purpose, the main discipline for safety design engineers is to perform a detailed Cause-Consequence Diagram (CCD)~\cite{andrews2001reliability} reliability analysis for identifying the subsystem events that prevent the entire system from functioning as desired. This approach models the causes of  component failures and their consequences on the entire system using Fault Tree~(FT)~\cite{towhidnejad2002fault} and Event Tree~(ET)~\cite{papazoglou1998mathematical} dependability~modeling~techniques.\\

FTs mainly provide a graphical model for analyzing the factors causing a system failure upon their occurrences. FTs are generally classified into two categories Static Fault Trees~(SFT) and Dynamic Fault Trees~(DFT) \cite{backstrom2016effective}. SFTs and DFTs allow safety-analysts to capture the static/dynamic failure characteristics of systems in a very effective manner using \textit{logic}-gates, such as OR, AND, NOT, Priority-AND~(PAND) and SPare (SP)~\cite{towhidnejad2002fault}. However, the FT technique is incapable of identifying the possible consequences resulting from an undesirable failure on the entire system.
ETs provide risk analysis with all possible system-level operating states that can occur in the system, i.e., success and failure, so that one of these possible scenarios can occur~\cite{papazoglou1998mathematical}. However, both of these modeling techniques are limited to analyzing either a critical-system failure or cascading dependencies of system-level components only, respectively. \\

There exist some techniques that have been developed for subsystem-level reliability analysis of safety-critical systems. For instance, Papadopoulos et al. in~\cite{papadopoulos2011engineering} have developed a software tool called \textit{HiP-HOPS} (Hierarchically Performed Hazard Origin \& Propagation Studies) \cite{hiphops_tp} for subsystem-level failure analysis to overcome classical manual failure analysis of complex systems and prevent human errors. HiP-HOPS can automatically generate the subsystem-level FT and perform Failure Modes, Effects, and Critically Analyses (FEMCA) from a given system model, where each system component is associated with its failure rate or failure probability~\cite{papadopoulos2011engineering}. Currently, HiP-HOPS lacks the modeling of \textit{multi-state} system components and also cannot provide generic mathematical expressions that can be used to predict the reliability of a critical-system based on any probabilistic distribution~\cite{kabir2019conceptual}. Similarly, Jahanian in~\cite{jahanian2019failure} has proposed a new technique called Failure Mode Reasoning~(FMR) for identifying and quantifying the failure modes for safety-critical systems at the subsystem level. However, according to Jahanian~\cite{jahanian2020failure}, the soundness of the FMR approach needs to be proven mathematically. \\

On the other hand, CCD analysis typically uses FTs to analyze failures at the subsystem or component level combined with an ET diagram to integrate their cascading failure dependencies at the system level. CCDs are categorized into two general methods for the ET linking process with the FTs~\cite{vcepin2011assessment}: (1) Small ET diagram and large subsystem-level FT; (2) Large ET diagram and small subsystem-level FT. The former one with small ET and large subsystem-level FT is the most commonly used for the probabilistic safety assessment of industrial applications (e.g., in \cite{liu2008probabilistic}). There are \textit{four} main steps involved in the CCD analysis~\cite{andrews2002application}: (1)~\textit{Component failure events}:~identify the causes of each component failure associated with their different modes of operations;  (2)~\textit{Construction of a complete CCD}:~construct a CCD model using its basic blocks, i.e., \textit{Decision box}, \textit{Consequence path} and \textit{Consequence box}; (3)~\textit{Reduction}: removal  of unnecessary decision boxes based on the system functional behavior to obtain a minimal CCD; and lastly (4)~\textit{Probabilistic analysis}:~evaluating the probabilities of CCD paths describing the occurrence of a sequence of~events. \\

Traditionally, CCD subsystem-level reliability analysis is carried out by using paper-and-pencil-based approaches to analyze safety-critical systems, such as high-integrity protection systems~(HIPS)~\cite{andrews2002application} and nuclear power plants~\cite{ridley2000dependency}, or using computer simulation tools based on Monte-Carlo approach, as  in~\cite{bevilacqua2000monte}.  A major limitation in both of the above approaches is the possibility of introducing inaccuracies in the CCD analysis either due to human infallibility or the approximation errors due to numerical methods and pseudo-random numbers in the simulation tools. Moreover, simulation tools do not provide the mathematical expressions that can be used to predict the reliability of a given system based on any probabilistic distributions~and~failure~rates. \\

A more safe  way is to substitute  the  error-prone  informal  reasoning  of  CCD analysis by  formal  generic mathematical proofs as per recommendations of safety standards, such as IEC 61850 \cite{mackiewicz2006overview}, EN 50128 \cite{gallina2016deriving} and ISO 26262 \cite{palin2011iso}. In this work, we propose to use formal techniques based on theorem proving for the formal reliability CCD analysis-based of safety-critical systems, which provides us the ability to obtain a \textit{verified} subsystem-level failure/operating consequence expression. Theorem proving is a formal verification technique~\cite{hasan2015formal}, which is used for conducting the proof of mathematical theorems based on a computerized proof tool.  In particular, we use HOL4 \cite{HOL_tp}, which is an interactive theorem prover with the ability of verifying a wide range of mathematical expressions constructed in higher-order logic~(HOL). For this purpose, we endeavor to formalize the above-mentioned \textit{four} steps of CCD analysis using HOL4 proof assistant. To demonstrate the practical effectiveness of the proposed CCD formalization, we conduct the formal CCD analysis of  an IEEE 39-bus electrical power network system. Subsequently, we formally determine a commonly used metric, namely Forced Outage Rate ($\mathcal{FOR}$), which determines the capacity outage or unavailability of the power generation units~\cite{grainger2003power}. Also, we evaluate the System Average Interruption Duration Index~($\mathcal{SAIDI}$), which describes the average duration of interruptions for each customer in a power network~\cite{grainger2003power}.\\

The main contributions of the work we describe in this report can be summarized as follows:

\begin{itemize}
  \item[$\bullet$] Formalization of the CCD basic constructors, such as  \textit{Decision box}, \textit{Consequence path} and \textit{Consequence box}, that can  be used to build an arbitrary level of~CCDs
  \item[$\bullet$] Enabling the formal reduction of CCDs that can remove unnecessary decision boxes from a given CCD model, a feature not available in other existing approaches
  \item[$\bullet$] Provide reasoning support for formal probabilistic analysis of scalable CCDs consequence paths with new proposed mathematical formulations
  \item[$\bullet$]  Application on a real-world IEEE 39-bus electrical power network system and verification of its reliability indexes $\mathcal{FOR}$ and $\mathcal{SAIDI}$
  \item[$\bullet$]  Development of a Standard Meta Language (SML) function that can numerically compute reliability values from the \textit{verified} expressions of $\mathcal{FOR}$ and $\mathcal{SAIDI}$
  \item[$\bullet$] Comparison between our formal CCD reliability assessment with the corresponding results obtained~from MATLAB MCS and other notorious approaches
\end{itemize}

The rest of the report is organized as follows: In Section~\ref{Related Work}, we present the related literature review. In Section~\ref{Preliminaries}, we describe the preliminaries to facilitate the understanding of the rest of the report. Section~\ref{Cause-Consequence Diagrams} presents the proposed formalization of CCD and its formal probabilistic properties. In Section~\ref{Electrical Power 39-bus Network System}, we describe the formal CCD analysis of an electrical network system and the evaluation of its reliability indices  $\mathcal{FOR}$ and $\mathcal{SAIDI}$. Lastly, Section~\ref{Conclusions} concludes~the~report.

\section{Related Work}
\label{Related Work}
Only a few work have previously considered using formal techniques~\cite{hasan2015formal} to model and analyze CCDs. For instance, Ortmeier et al. in~\cite{ortmeier2005deductive} developed a framework for Deductive Cause-Consequence Analysis (DCCA) using the SMV model checker \cite{SMV_tp} to verify the CCD proof obligations. However, according to the authors~\cite{ortmeier2005deductive}, there is a problem of showing the completeness of DCCA due to the exponential growth of the number of proof obligations with complex systems that need cumbersome proof efforts. To overcome above-mentioned limitations, a more practical way is to verify \textit{generic} mathematical formulations that can perform $\mathcal{N}$-level CCD reliability analysis for real-world systems within a sound environment. Higher-Order-Logic (HOL) \cite{miller2012programming} is a good candidate formalism for achieving this goal. \\

Prior to our work, there were two notable projects for building frameworks to formally analyze dependability models using HOL4 theorem proving \cite{HOL_tp}. For instance, HOL4 has been previously used by Ahmad et al. in \cite{ahmad2015} to formalize SFTs. The SFT formalization includes a new datatype consisting of \texttt{AND}, \texttt{OR} and \texttt{NOT} FT gates~\cite{towhidnejad2002fault} to analyze the factors causing a static system failure. Furthermore, Elderhalli et al. in \cite{elderhalli2019methodology} had formalized DFTs in the HOL4 theorem prover, which can be used to conduct formal dynamic failure analysis. Similarly, we have defined in~\cite{Etree_tp1}  a new \texttt{EVENT\_TREE} datatype to model and analyze all possible system-level success and failure relationships. All these formalizations are basically required to formally analyze either a system static/dynamic failure or cascading dependencies of system-level components only, respectively. On the other hand, CCDs have the superior capability to use SFTs/DFTs for analyzing the static/dynamic failures at the subsystem level and analyze their cascading dependencies at the system-level using ETs. For that purpose, in this work, we provide new formulations that can model mathematically the graphical diagrams of CCDs and perform the subsystem-level reliability analysis of  highly-critical systems. Moreover, our proposed new mathematics provides the modeling of \textit{multi-state} system components and is based on any given probabilistic distribution and failure rates, which makes our proposed work the first of its kind. In order to check the correctness of the proposed equations, we verified them within the sound environment of  HOL4.
\section{Preliminaries} 
\label{Preliminaries}
In this section, we  briefly summarize the fundamentals of the HOL4 theorem proving approach and existing FT and ET formalizations in HOL4 to facilitate the reader's understanding of the rest of the~report. 

\subsection{HOL4 Theorem Proving}
\label{HOL4 Theorem Proving}

Theorem proving  \cite{hasan2015formal} is one of the formal verification techniques that use a computerized proof system for conducting the proof of mathematical theorems. HOL4 \cite{HOL_tp} is an interactive theorem prover, which is capable of verifying a wide range of safety-critical systems as well as mathematical expressions constructed in HOL. In general, given a safety-critical system to be formally analyzed, we first model its structure mathematically, then using the HOL4 theorem prover,  several properties of the system can be verified based on this mathematical model. The main characteristic of the HOL4 theorem prover is that its core consists only of four axioms and eight inference rules. Any further proof or theorem should be formally verified based on these axioms and rules or based on previously proven theorems. This ensured the soundness of the system model analysis, i.e., no wrong proof goal can be proved. Moreover, since the system properties are proven mathematically within HOL4, no approximation is involved in the analysis results. These features make HOL4 suitable for carrying out the CCD-based reliability analysis of safety-critical systems that require \textit{sound verification} results. Table \ref{Table: HOL4 Symbols and Functions} provides the HOL4 symbols and functions that we will use in this report. 

\begin{table} [!h]
 \caption{\protect  HOL4 Symbols and Functions}
 \label{Table: HOL4 Symbols and Functions}
\begin{center}
 \begin{tabular}{|l|l|l|} 
 \hline \thead{\makecell[l]{HOL4 Symbol}} &  \thead{\makecell[l]{Standard}} &  \thead{\makecell[l]{Meaning}} \\ 
 \hline  \hline
 \makecell[l]{\{x $|$ P(x)\}} & \makecell[l]{\{$\lambda x$. $P(x)$\}}  &  \makecell[l]{Set of all $x$ such that $P(x)$}  \\ [0.5ex] 
 \hline
 \makecell[l]{h :: L} & \makecell[l]{$cons$}  &  \makecell[l]{Add an element $h$ to a list L}  \\ [0.5ex] 
 \hline
  \makecell[l]{MAP ($\lambda$x. f(x)) X} & \makecell[l]{x $\in$ X $\to$ ($\lambda$x. f)}  &  \makecell[l]{Function that maps each \\ element \textit{x}   in the list X to f(x)}  \\ [0.5ex] 
 \hline
  \makecell[l]{$\mathrm{L}_1$ $++$ $\mathrm{L}_2$ } & \makecell[l]{$append$}  &  \makecell[l]{ Joins lists $\mathrm{L}_1$ and $\mathrm{L}_2$ together}  \\  [0.5ex] 
 \hline

\end{tabular}
\end{center}
\end{table}

\subsection{Probability Theory in HOL4}
\label{Probability Theory in HOL4}

Measure space is defined mathematically as ($\Omega$, $\Sigma$,~and~$\mu$), where $\Omega$ represents the sample space, $\Sigma$ represents
a $\sigma$-algebra of subsets of $\Omega$, and $\mu$ represents a measure with the domain~$\Sigma$. A probability space is a measure space ($\Omega$,~$\Sigma$,~and~Pr), where $\Omega$ is the complete sample space, $\Sigma$ is the corresponding event space containing all the events of interest, and Pr is the probability measure of the sample space as 1. The HOL4 theorem prover has a rich library of probabilities, including the functions \texttt{p\_space}, \texttt{events}, and \texttt{prob}. Given a probability space~\textit{p}, these functions return the corresponding $\Omega$, $\Sigma$, and Pr, respectively. The Cumulative Distribution Function~(CDF) is defined as the probability of the event where a random variable \textit{X} has a value less or equal to a value~\textit{t}, i.e.,  $\mathcal{P} (X \leq t)$. This definition can be been formalized in HOL4 as~\cite{hasan2009formal}:
\begin{flushleft}
	\vspace{1pt} \small{\texttt{$\vdash$ CDF p X t =  distribution p X \{y | y $\leq$ t\}}}
\end{flushleft}

\noindent where the function \texttt{distribution} takes three inputs: (i)~probability space~\textit{p}; (ii) random variable \textit{X}; and (iii)~set of real numbers, then returns the probability of the variable \textit{X} acquiring all the values of the given set in probability~space~\textit{p}. \\

\subsection{FT Formalization}
Fault Tree (FT) analysis \cite{towhidnejad2002fault} is one of the commonly used reliability assessment techniques for critical-systems. It mainly provides a schematic diagram for analyzing undesired \textit{top events}, which can cause complete system failure upon their occurrence. An FT model is represented by \textit{logic}-gates, like OR, AND and NOT, where an OR gate models the failure of the output if any  of  the input failure events occurs alone, while an AND gate models the failure of the output if all of the input failure events occur at the same time, and lastly a NOT gate models the complement of the input failure event. Ahmad et al.~\cite{ahmad2015} presented the FT formalization by defining a new datatype gate, in HOL4~as:

\begin{flushleft}
	\vspace{1pt} \small{ \texttt{\bf{Hol\_datatype}}  
		\texttt{gate =  AND of (gate list) | \\ \qquad  \qquad  \qquad \qquad \hspace{7mm} OR of (gate list) |  \\ \qquad  \qquad  \qquad \qquad \hspace{7mm}  NOT of (gate) |   \\ \qquad  \qquad  \qquad \qquad \hspace{7mm} atomic of (event)}}
\end{flushleft}

\noindent The FT constructors \texttt{AND} and \texttt{OR} are recursive functions on \texttt{gate}-typed lists, while the FT constructor \texttt{NOT} operates on a \texttt{gate}-type variable. A semantic function is then defined over the \texttt{gate} datatype that can yield an FT diagram~as:\\

\begin{flushleft}
	\texttt{\bf{Definition 1:}}\\
	\vspace{1pt} 
	\small{\texttt{$\vdash$  FTree p (atomic X) = X}} $\wedge$	\\ \hspace{3mm}
	\small{\texttt{FTree p (OR \hspace{0.1mm} (h::t))  =  FTree p h $\cup$ FTree p (OR \hspace{0.01mm} t)}} $\wedge$ \\ \hspace{3mm}  
	\small{\texttt{FTree p (AND (h::t)) = FTree p h $\cap$ FTree p (AND t)}} $\wedge$ \\ \hspace{3mm}
	\small{\texttt{FTree p (NOT X) = p\_space p DIFF FTree p X}}  
\end{flushleft}

\noindent The function \texttt{FTree} takes an event \texttt{X}, identified by a type constructor \texttt{atomic}, and returns the given event \texttt{X}. If the function \texttt{FTree} takes a list of type \texttt{gate}, identified by a type constructor \texttt{OR}, then it returns the union of all elements after applying the function \texttt{FTree} on each element of the given list. Similarly, if the function \texttt{FTree} takes a list of type \texttt{gate}, identified by a type constructor \texttt{AND}, then it performs the intersection of all elements after applying the function \texttt{FTree} on each element of the given list. For the \texttt{NOT} type constructor, the function \texttt{FTree} returns the complement of the failure event obtained from the function~\texttt{FTree}. \\

\begin{table} [!b]
 \caption{\protect  FT HOL4 Probabilistic Theorems}
 \label{Table: Probability of Failures of Fault Tree Gates}
\begin{center}
 \begin{tabular}{|c|l|} 
 \hline \thead{FT Gate} & \thead{Probabilistic Theorem} \\ [0.5ex] 
 \hline \hline
  \makecell{\includegraphics[width=0.3\textwidth]{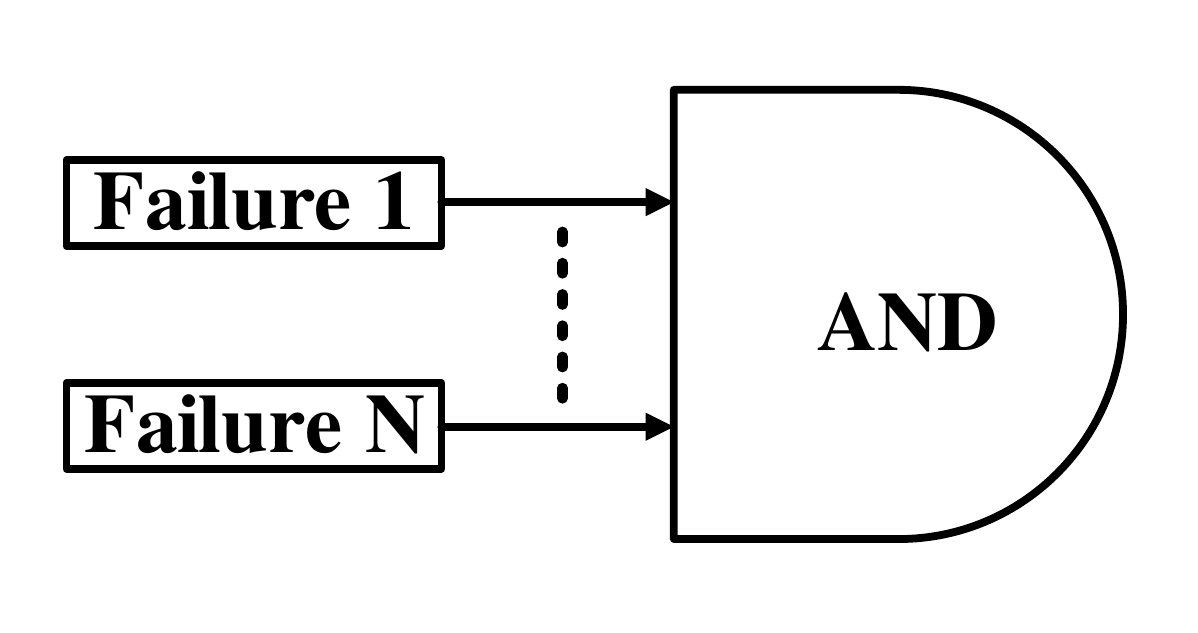}}  & \makecell[l]{\small{\texttt{prob p}} \\ \small{\texttt{\hspace{0.1mm}  (FTree p (AND $\mathrm{F}_\mathcal{N}$)) = $\prod$ (PROB\_LIST p $\mathrm{F}_\mathcal{N}$)}}}  \\  [0.5ex]
 \hline
  \makecell{\includegraphics[width=0.3\textwidth]{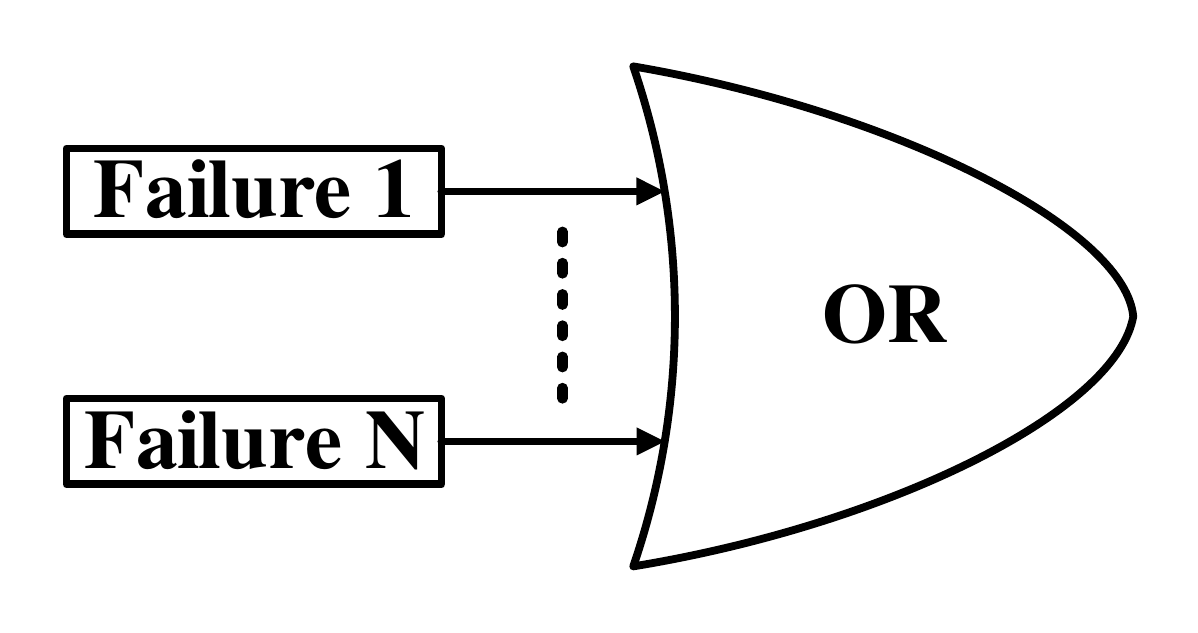}}  & \makecell[l]{\small{\texttt{prob p }} \\ \small{\texttt{\hspace{1.5mm}  (FTree p (OR $\mathrm{F}_\mathcal{N}$)) = }} \\  \hspace{10mm} \small{\texttt{1 - $\prod$ (PROB\_LIST p (COMPL\_LIST p $\mathrm{F}_\mathcal{N}$))}}}  \\  [0.5ex]
 \hline 
\end{tabular}
\end{center}
\end{table} 

The formal verification in HOL4 for the failure probabilistic expressions of the above-mentioned FT gates is presented in Table \ref{Table: Probability of Failures of Fault Tree Gates} \cite{ahmad2015}.  These expressions are verified under the following constrains: (a) \texttt{$F_\mathcal{N}$ $\in$ events p} ensures that all associated failure events in the given list $F_\mathcal{N}$ are drawn from the events space~\textit{p}; (b) \texttt{prob\_space~p} ensures that \textit{p} is a valid probability space; and lastly (c)~\texttt{MUTUAL\_INDEP~p~$F_\mathcal{N}$} ensures the independence of the failure events in the given list $F_\mathcal{N}$. The function $\prod$  takes a list and returns the product of the list elements while the function \texttt{PROB\_LIST} returns a list of probabilities associated with the elements of the list. The function \texttt{COMPL\_LIST} returns the complement of the given list elements.

\subsection{ET Formalization}
Event Tree (ET) \cite{papazoglou1998mathematical} analysis is a widely used technique to enumerate all possible combinations of system-level components failure and success states in the form of a tree structure. An ET diagram starts by an initiating event called~\textit{Node} and then all possible scenarios of an event that can occur in the system are drawn as \textit{Branches}. ETs were formally modeled by using a new recursive datatype \texttt{EVENT\_TREE}, in HOL4~as~\cite{Etree_tp1}:
\begin{flushleft}
	\vspace{1pt} \small{ \texttt{\bf{Hol\_datatype}}  
		\texttt{EVENT\_TREE =  ATOMIC of (event) |  \\ \qquad \qquad \qquad \qquad \qquad \qquad \hspace{2mm}  NODE of  (EVENT\_TREE list) | \\ \qquad \qquad \qquad \qquad \qquad \qquad \hspace{2mm} BRANCH of (event) (EVENT\_TREE list)}}
\end{flushleft}
\noindent The type constructors \texttt{NODE} and \texttt{BRANCH} are recursive functions on \texttt{EVENT\_TREE}-typed lists. A semantic function is then defined over the \texttt{EVENT\_TREE} datatype that can yield a corresponding ET diagram as:
\begin{flushleft}
\label{Definition 6}
	\texttt{\bf{Definition 2:}}\\
	\vspace{1pt} 
	\small{\texttt{$\vdash$ ETREE (ATOMIC X) = X $\wedge$}} \\ 
	\small{\texttt{\hspace{2mm} ETREE (NODE (h::L)) =  ETREE h $\cup$ (ETREE (NODE L))}}  $\wedge$ \\    	
	\small{\texttt{\hspace{2mm} ETREE (BRANCH X (h::L)) =  X $\cap$ (ETREE h $\cup$ ETREE (BRANCH X L))}}    	
\end{flushleft}

\begin{table} [!b]
 \caption{\protect  ET HOL4 Probabilistic Theorems}
 \label{Table: HOL probabilistic theorems of ET constructors}
\begin{center}
 \begin{tabular}{|l|l|} 
 \hline \thead{\makecell{ET Constructor}} & \thead{Probabilistic Theorem} \\ 
 \hline \hline
  \makecell{\includegraphics[width=0.2\columnwidth]{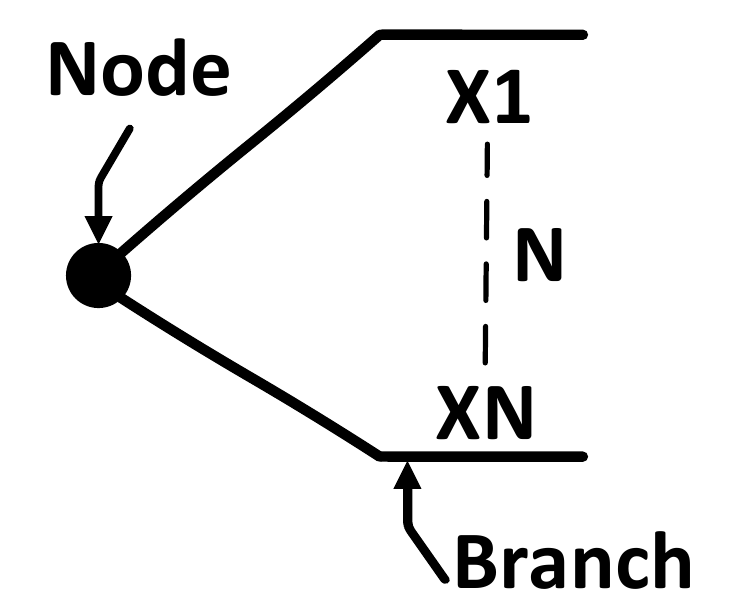}} &  \makecell[l]{\small{\texttt{prob p  (ETREE (NODE $\mathcal{X}_\mathcal{N}$)) =}}    \small{\texttt{$\sum_{\mathcal{P}}$ p $\mathcal{X}_\mathcal{N}$}}}  \\  
 \hline
 \makecell{\includegraphics[width=0.2\columnwidth]{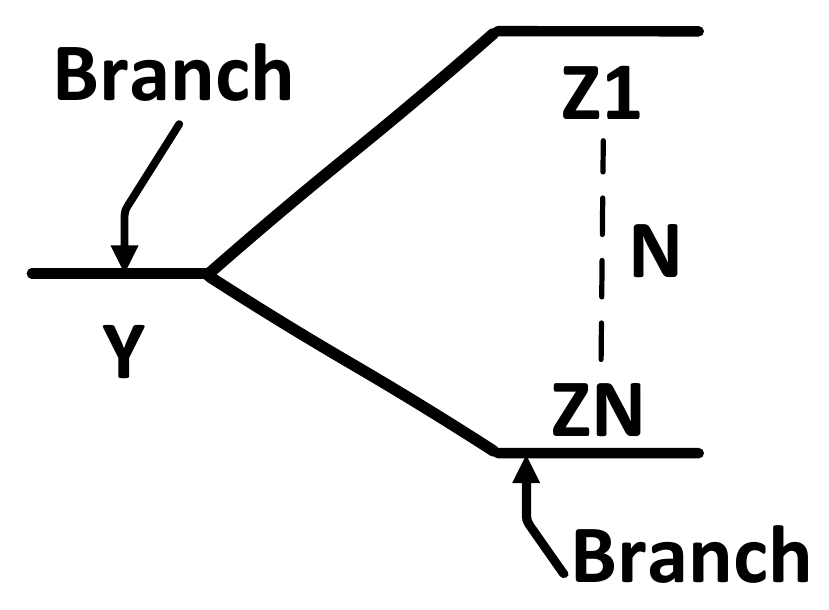}} &  \makecell[l]{\small{\texttt{prob p }} \\ \small{\texttt{\hspace{1mm} (ETREE (BRANCH Y $\mathcal{Z}_\mathcal{N}$))  = (prob p Y) $\times$  $\sum_{\mathcal{P}}$ p $\mathcal{Z}_\mathcal{N}$}}}  \\ 
 \hline 
\end{tabular}
\end{center}
\end{table}

\noindent The function \texttt{ETREE} takes an event \texttt{X}, identified by a type constructor \texttt{ATOMIC} and returns the event \texttt{X}. If the function \texttt{ETREE} takes a list of type \texttt{EVENT\_TREE}, identified by a type constructor \texttt{NODE}, then it returns the union of all elements after applying the function \texttt{ETREE} on each element of the list. Similarly, if the function \texttt{ETREE} takes an event \texttt{X} and a list of type \texttt{EVENT\_TREE}, identified by a type constructor \texttt{BRANCH}, then it performs the intersection of the event \texttt{X}  with the union of the head of the list after applying the function \texttt{ETREE} and the recursive call of the \texttt{BRANCH}~constructor. For the formal probabilistic assessment of each path occurrence in the ET diagram, HOL4 probabilistic properties for \texttt{NODE} and \texttt{BRANCH} ET constructors are presented in Table~\ref{Table: HOL probabilistic theorems of ET constructors}~\cite{Etree_tp1}.  These expressions are formally verified under the same FT constrains, i.e.,  \texttt{$\mathcal{X}_\mathcal{N}$ $\in$ events p}, \texttt{prob\_space~p} and \texttt{MUTUAL\_INDEP~p~$\mathcal{X}_\mathcal{N}$}. The function $\sum_{\mathcal{P}}$ is defined to sum the probabilities of events for a list. 

\section{Cause-Consequence Diagrams}
\label{Cause-Consequence Diagrams}
\begin{figure}[H]
	\includegraphics[width= 0.75 \columnwidth]{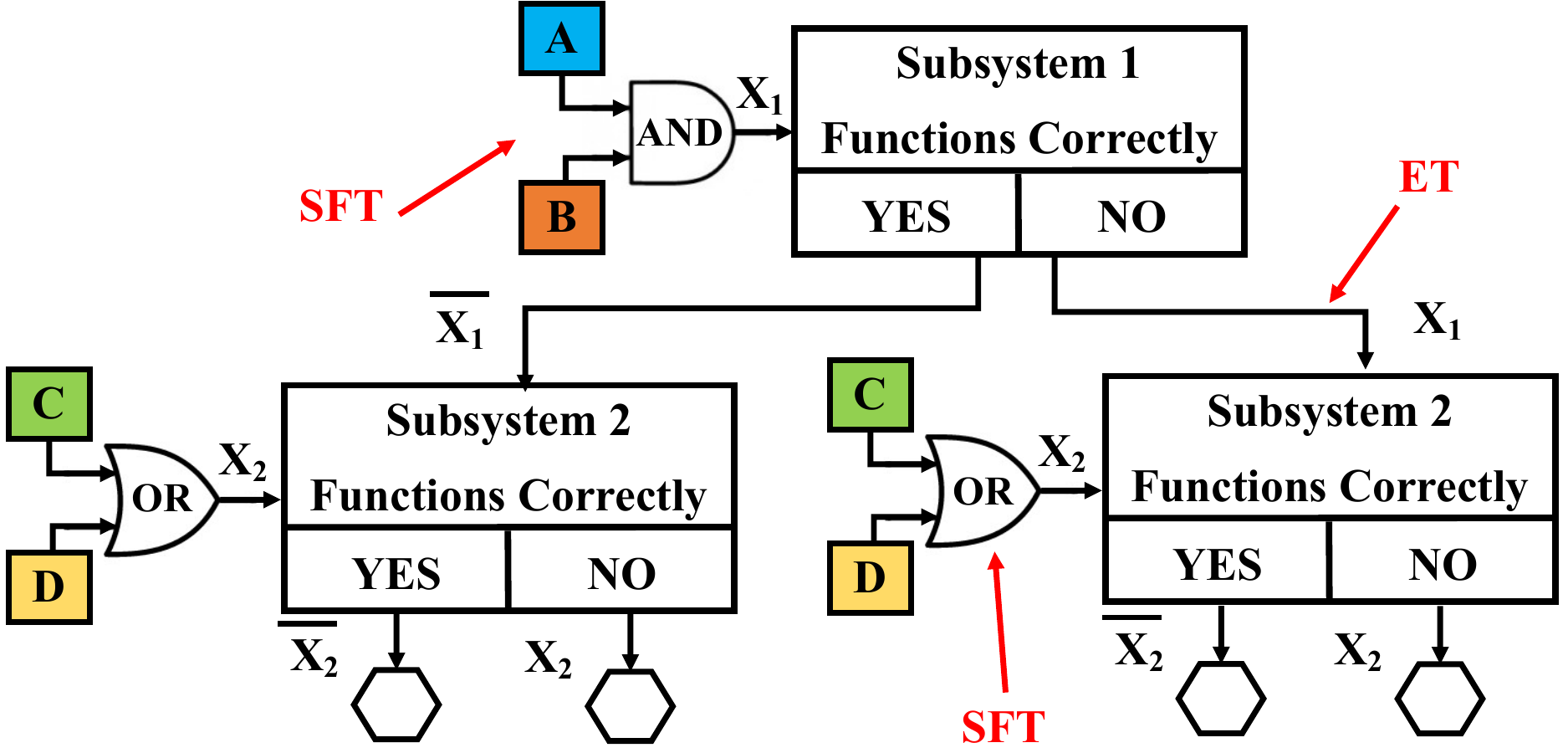} 
	\centering
	\caption{\protect  CCD Analysis Type A}
	\label{Fig: CCD analysis Type A}
\end{figure}

\begin{figure}[H]
	\includegraphics[width= 0.75 \columnwidth]{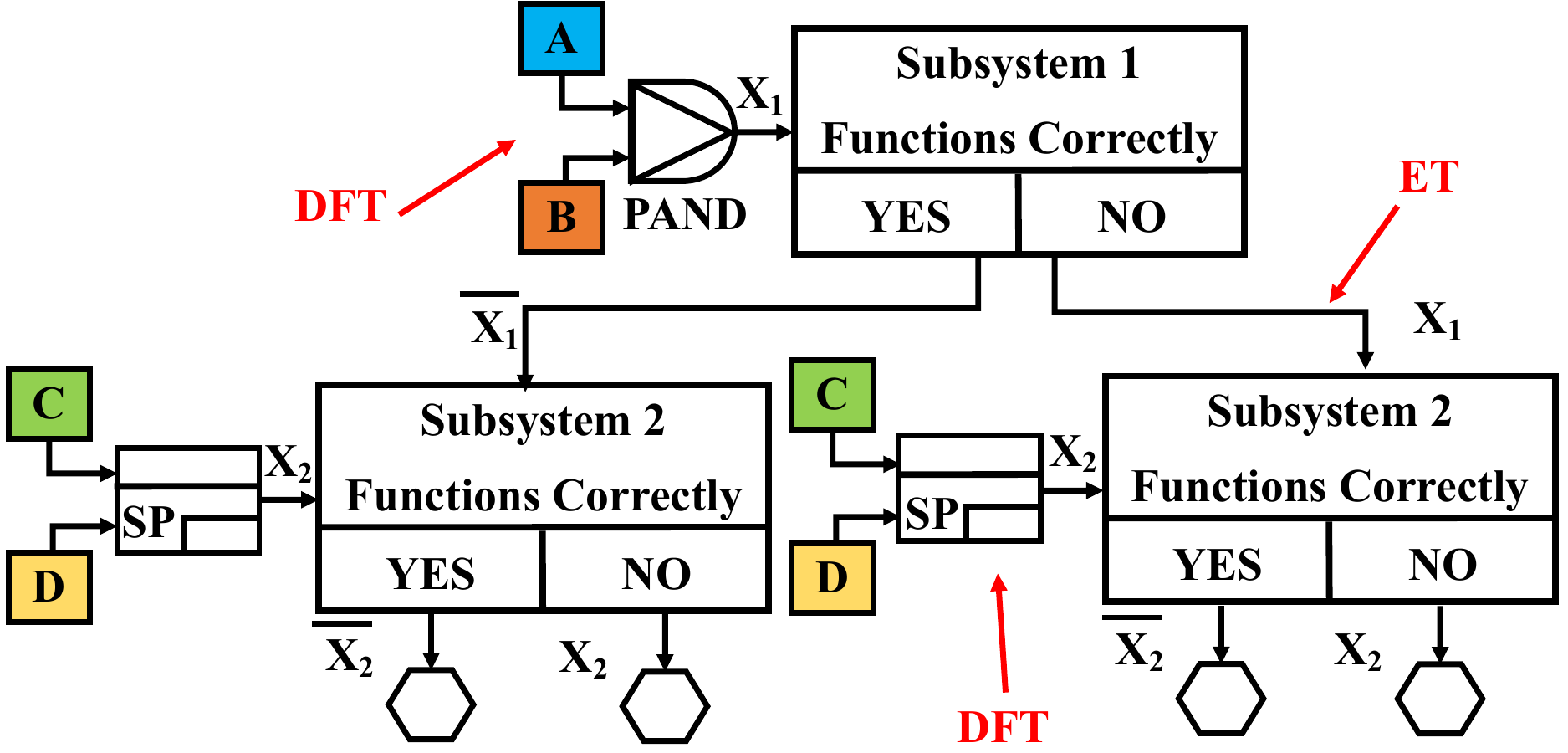} 
	\centering
	\caption{\protect  CCD Analysis Type B}
	\label{Fig: CCD analysis Type B}
\end{figure}

Cause–Consequence Diagram \cite{ridley2000dependency} (CCD) has been developed to analyze the causes of an undesired subsystem failure events, using FT analysis, and from these events obtain all possible consequences on the entire system, using  ET analysis~\cite{vyzaite2006cause}. The description of the CCD basic constructors are illustrated in Table \ref{Table: CCD Symbol} \cite{andrews2002application}. CCD analysis is mainly divided into two  categories~\cite{xu2004combining}: (1) \textit{Type~I} that combines SFT and ET, as shown in Fig.~\ref{Fig: CCD analysis Type A} and Table \ref{Table: SFT Symbols and Functions} \cite{vcepin2011assessment}; and (2)~\textit{Type~II} that combines DFT and ET without shared events in different subsystems, as shown in Fig.~\ref{Fig: CCD analysis Type B} and Table \ref{Table: DFT Symbols and Functions} \cite{vcepin2011assessment}. In this analysis, we focus on the CCD-based reliability analysis at the subsystem level of~\textit{Type I}. 

\begin{table} [!h]
 \caption{\protect  CCD Symbols and Functions}
 \label{Table: CCD Symbol}
\begin{center}
 \begin{tabular}{|c|l|} 
 \hline \thead{CCD Symbol} & \thead{Function} \\ [0.5ex] 
 \hline \hline
  \makecell{\includegraphics[width=0.35 \textwidth]{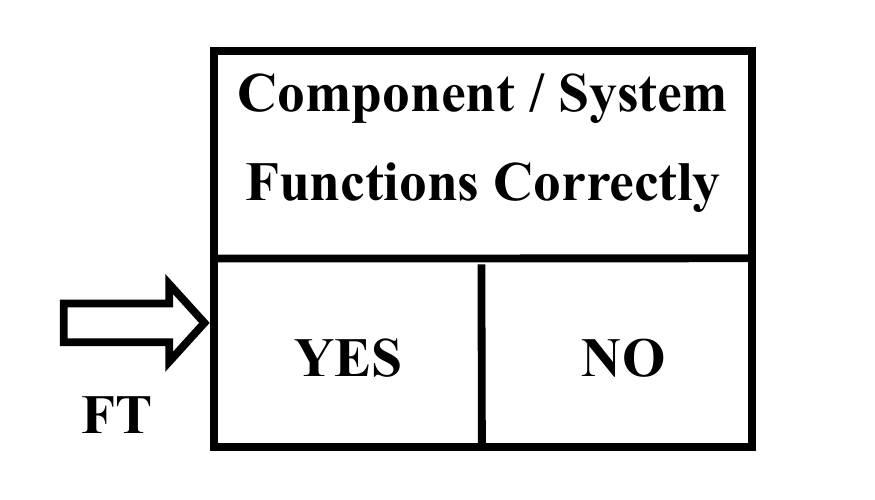}}  & \makecell[l]{\texttt{Decision Box:} represents the \\ functionality of a component. \\
  (1) \texttt{NO Box}: describes the component or \\ subsystem failure behavior.  A FT of the \\ component  is connected to this box  that \\ can be used  to determine the  failure \\ probability ($\mathcal{P}_F$) \\
  (2) \texttt{YES Box}: represents the correct \\ functioning of the  component or reliability, \\ which can be calculated by simply taking \\  the complement of the failure probability  \\ determined in the NO Box,   i.e., 1 - $\mathcal{P}_F$}  \\  [0.5ex]
 \hline
  \makecell{\includegraphics[width=0.15\textwidth]{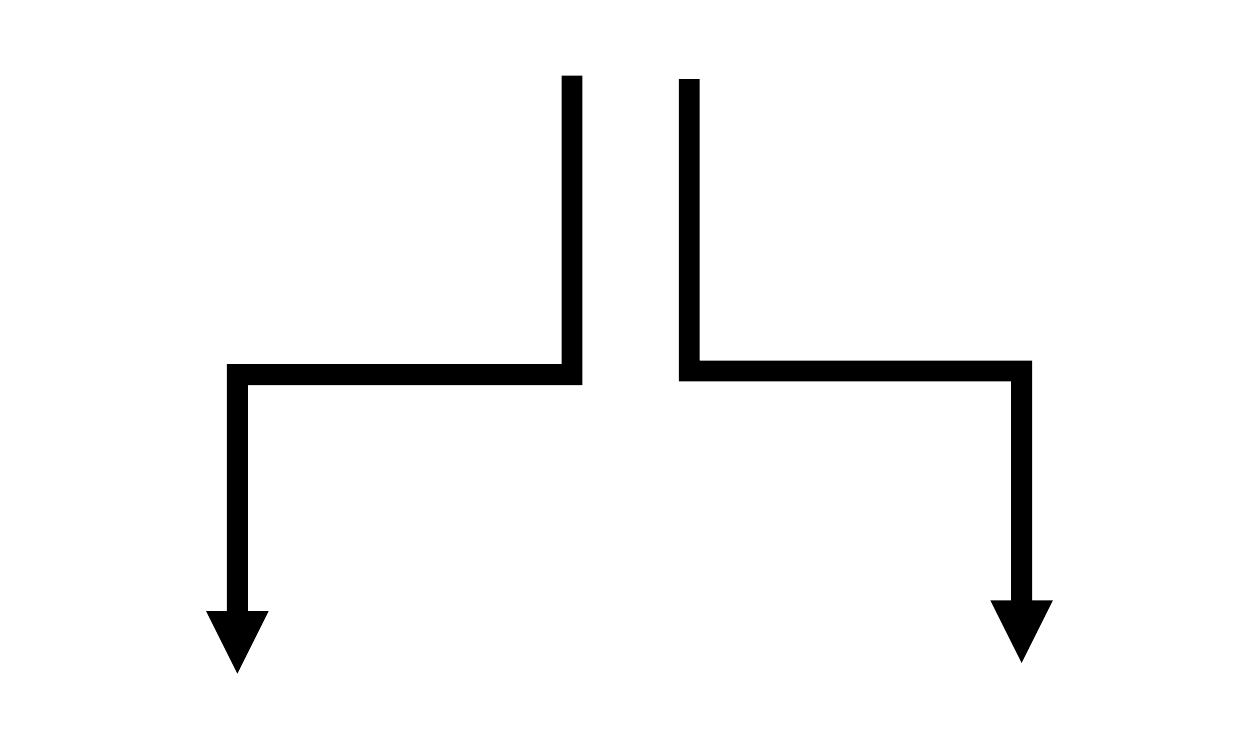}}  & \makecell[l]{\texttt{Consequence Path:} models the \\ next possible consequence scenarios due to \\ a particular event} \\ [0.5ex]
 \hline 
  \makecell{\includegraphics[width=0.15\textwidth]{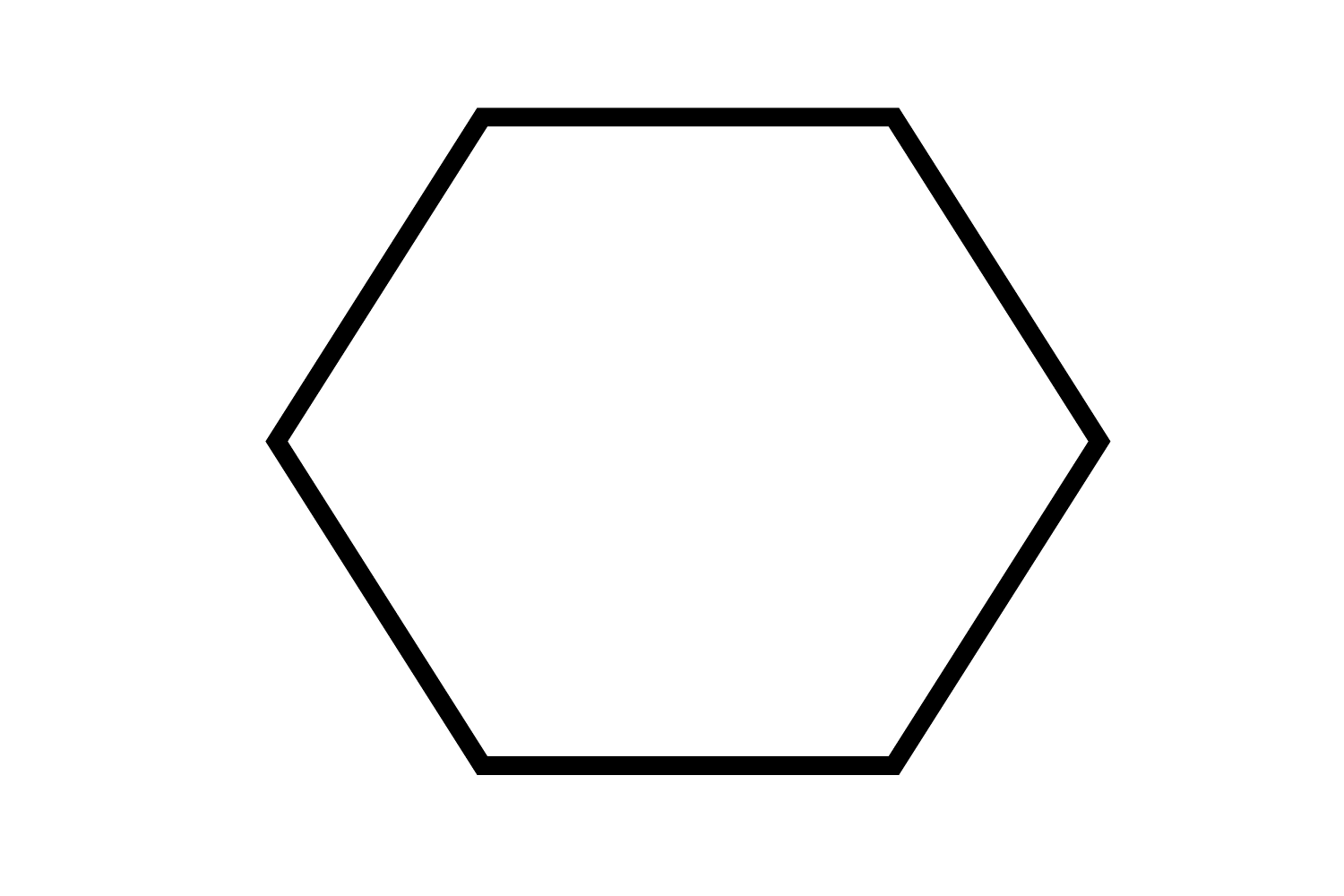}}  & \makecell[l]{\texttt{Consequence Box:}  models the \\ outcome event due  to a particular sequence \\  of events} \\ [0.5ex]
 \hline
\end{tabular}
\end{center}
\end{table}

\begin{table} [H]
 \caption{\protect  SFT Symbols and Functions}
 \label{Table: SFT Symbols and Functions}
\begin{center}
 \begin{tabular}{|l|l|} 
 \hline \thead{SFT Symbol} & \thead{Function} \\ [0.5ex] 
 \hline \hline
\makecell{\includegraphics[width=0.15\textwidth]{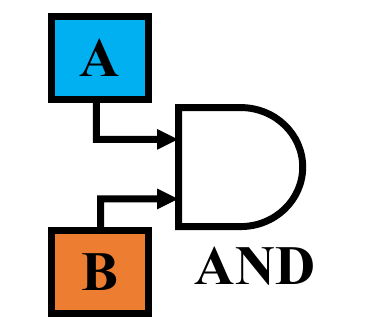}}  & \makecell[l]{\texttt{AND Gate:}  models the failure of the output if all \\
of the input failure events, i.e., A and B, occur at the \\ same time (simultaneously)} \\ [0.5ex]
\hline
 \makecell{\includegraphics[width=0.15\textwidth]{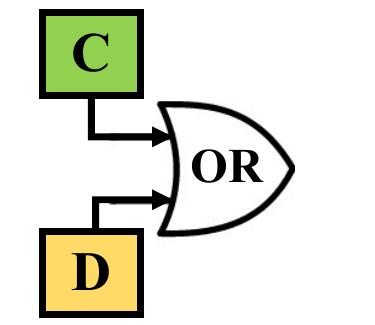}}  & \makecell[l]{\texttt{OR Gate:}  models the failure of the output if any \\ 
of the input failure events, i.e., C or D, occurs alone} \\ [0.5ex]
 \hline
\end{tabular}
\end{center}
\end{table}

\begin{table}[H]
 \caption{\protect  DFT Symbols and Functions}
 \label{Table: DFT Symbols and Functions}
\begin{center}
 \begin{tabular}{|l|l|} 
 \hline \thead{DFT Symbol} & \thead{Function} \\ [0.5ex] 
 \hline \hline
    \makecell{\includegraphics[width=0.15\textwidth]{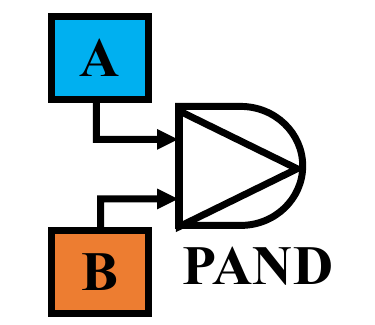}}  & \makecell[l]{\texttt{Priority-AND (PAND) Gate:}  models the \\ dynamic behavior of failing the output  when all \\  input events occur in a sequence, i.e., A then B} \\ [0.5ex]
 \hline
   \makecell{\includegraphics[width=0.15\textwidth]{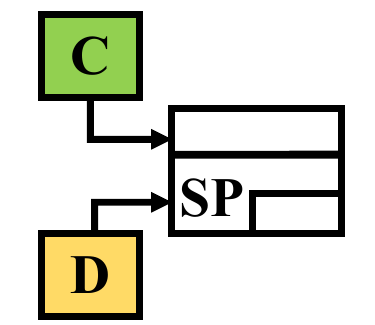}}  & \makecell[l]{\texttt{SPare (SP) Gate:}  models the dynamic behavior \\ of activating the spare input D after the failure of the \\ main input C} \\ [0.5ex]
 \hline

\end{tabular}
\end{center}
\end{table}

\begin{figure}[!b]
	\includegraphics[width= 1 \columnwidth]{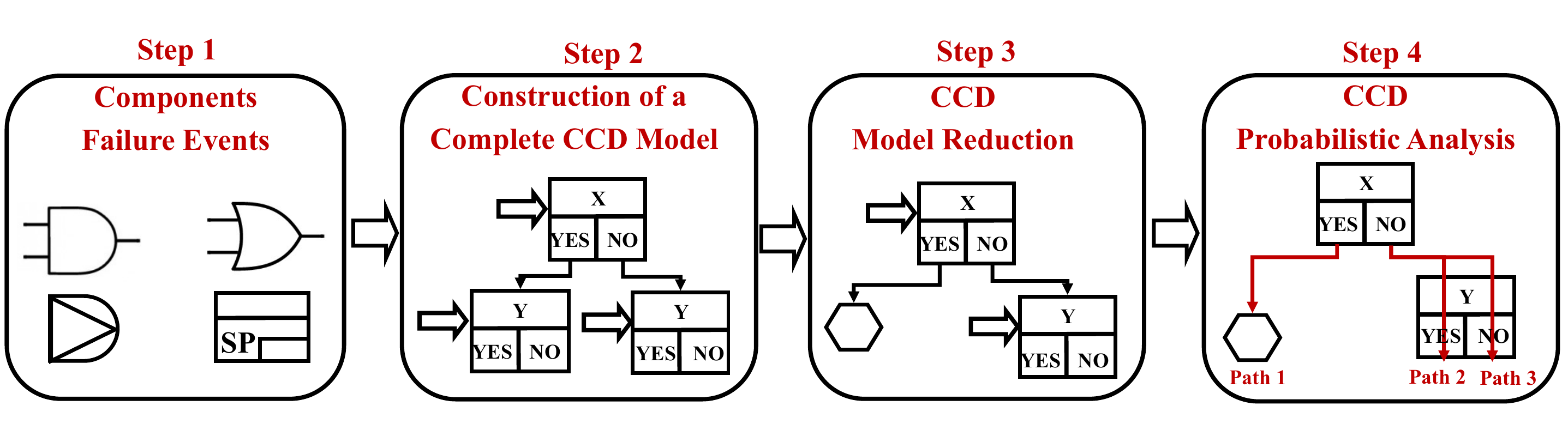} 
	\centering
	\caption{\protect  Overview of  CCD Analysis}
	\label{Fig: CCD}
\end{figure}

Fig. \ref{Fig: CCD} depicts the overview of the \textit{four} steps of CCD analysis~\cite{andrews2001reliability}: (1)~\textit{Components failure events}:~identify the causes of the undesired failure events for each subsystem/component in the safety-critical system;  (2)~\textit{Construction of a complete CCD}:~draw a complete system CCD model using its basic constructors considering that the order of components should follow the temporal action of the system; (3)~\textit{CCD model reduction}: remove the unnecessary decision boxes in the system to obtain its minimal CCD representing the actual functional behavior of the system; and (4)~\textit{CCD probabilistic analysis}:~evaluate the probabilities of all CCD consequence paths. The paths in a CCD represent the likelihood of specific sequence scenarios that are possible to occur in a system so that \textit{only one} scenario can occur~\cite{vyzaite2006cause}. This implies that all consequences in a CCD are disjoint (mutually exclusive)~\cite{andrews2002application}. Assuming that all events associated with the decision boxes in a CCD model are mutually independent, then the CCD paths probabilities can be quantified as~follows~\cite{ridley2000dependency}:

\begin{enumerate}
  \item Evaluate the probabilities of each outgoing branch stemming from a \textit{decision box}, i.e., quantifying the associated FT models
  \item Compute the probability of each \textit{consequence path} by multiplying the individual probabilities of all events associated with the decision boxes
  \item Determine the probability of a particular \textit{consequence box} by summing the probabilities of all consequence paths ending with that consequence event
\end{enumerate}

As an example, consider a Motor Control Centre~(MCC)~\cite{olsen2015enhanced} consisting of three components \textit{Relay}, \textit{Timer} and \textit{Fuse}, as shown in Fig.~\ref{Fig: MCC}. The MCC is designed to control an Induction Motor~(IM) and let it run for a specific period of time then stops. The IM power circuit is energized by the closure of the Relay Contacts (Rc), as shown in Fig.~\ref{Fig: MCC}. Rc closes after the user press the Start button that energizes R and at the same time energizes an ON-delay Timer~(T). The Timer opens its contacts~(Tc) after a specific period of time \textit{t} and consequently the IM stops. If the IM is overloaded than its design, then the Fuse (F) melts and protects both MCC and IM from damage. Assume that each component in the MCC has two operational states, i.e., operating or failing. The \textit{four} steps of a CCD-based reliability analysis described by Andrews et al.~\cite{andrews2002application} are as follows~\cite{vyzaite2006cause}:

\begin{figure}[!h]
	\includegraphics[width= 1 \columnwidth]{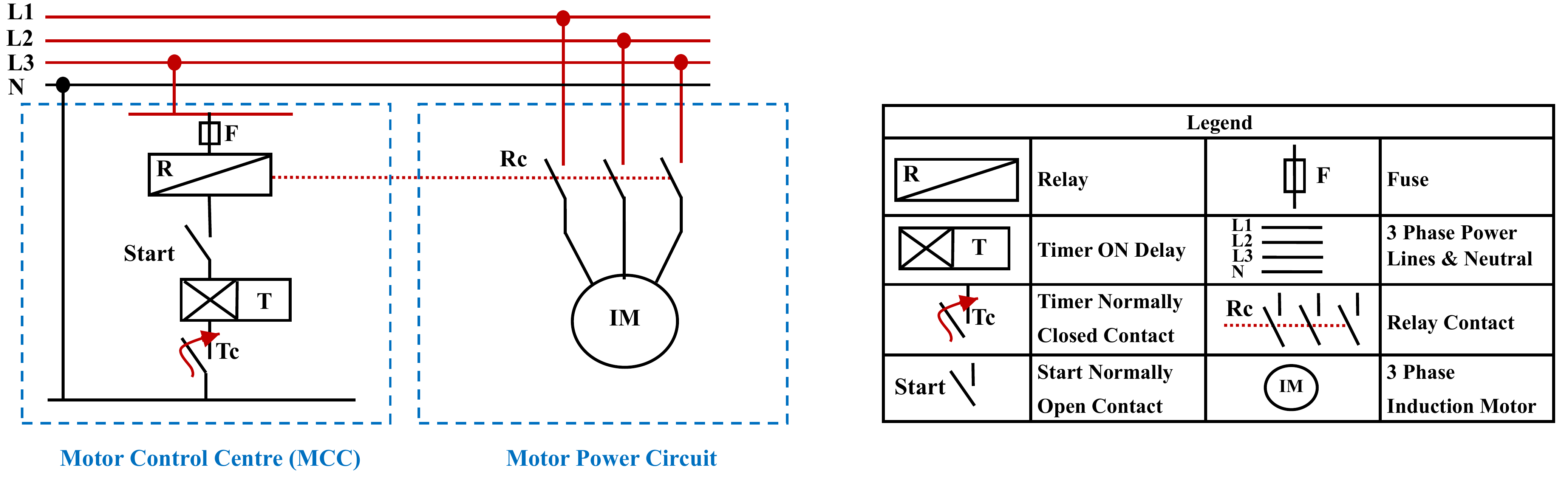} 
	\centering
	\caption{\protect   Schematic of an Example MCC}
	\label{Fig: MCC}
\end{figure}

\begin{enumerate}
\item \textit{Components failure events}: Assign a FT to each component in the MCC, i.e., $\mathrm{FT}_{Relay}$, $\mathrm{FT}_{Timer}$, $\mathrm{FT}_{Fuse}$.

\item \textit{Construction of a complete CCD}: Construct a complete CCD model  of the IM control operation,  as shown in Fig. \ref{Fig: Complete CCD Model}.  For instance,  if the condition of the first decision box  is either satisfied or not, i.e., YES or NO, then the next system components are considered in order, i.e., \textit{Timer} and \textit{Fuse}, respectively. Each consequence in the CCD ends with either motor stops (MS) or motor~runs~(MR). 

\item \textit{CCD model reduction}: Apply the reduction process on the obtained complete CCD model. For instance, if the condition of the first decision box  (Relay Contacts Open) is satisfied, i.e., YES box, then the IM stops regardless of the status of the rest of the components, as shown in Fig.~\ref{Fig:  Reduced CCD Model}. Similarly,  if the condition of the second decision box  (Timer Contacts Open) is satisfied, then the IM stops. So, Fig. \ref{Fig:  Reduced CCD Model} represents the minimal CCD for the IM control operation.\\

\begin{figure}[!h]
  \centering
  \includegraphics[width= 0.67 \columnwidth]{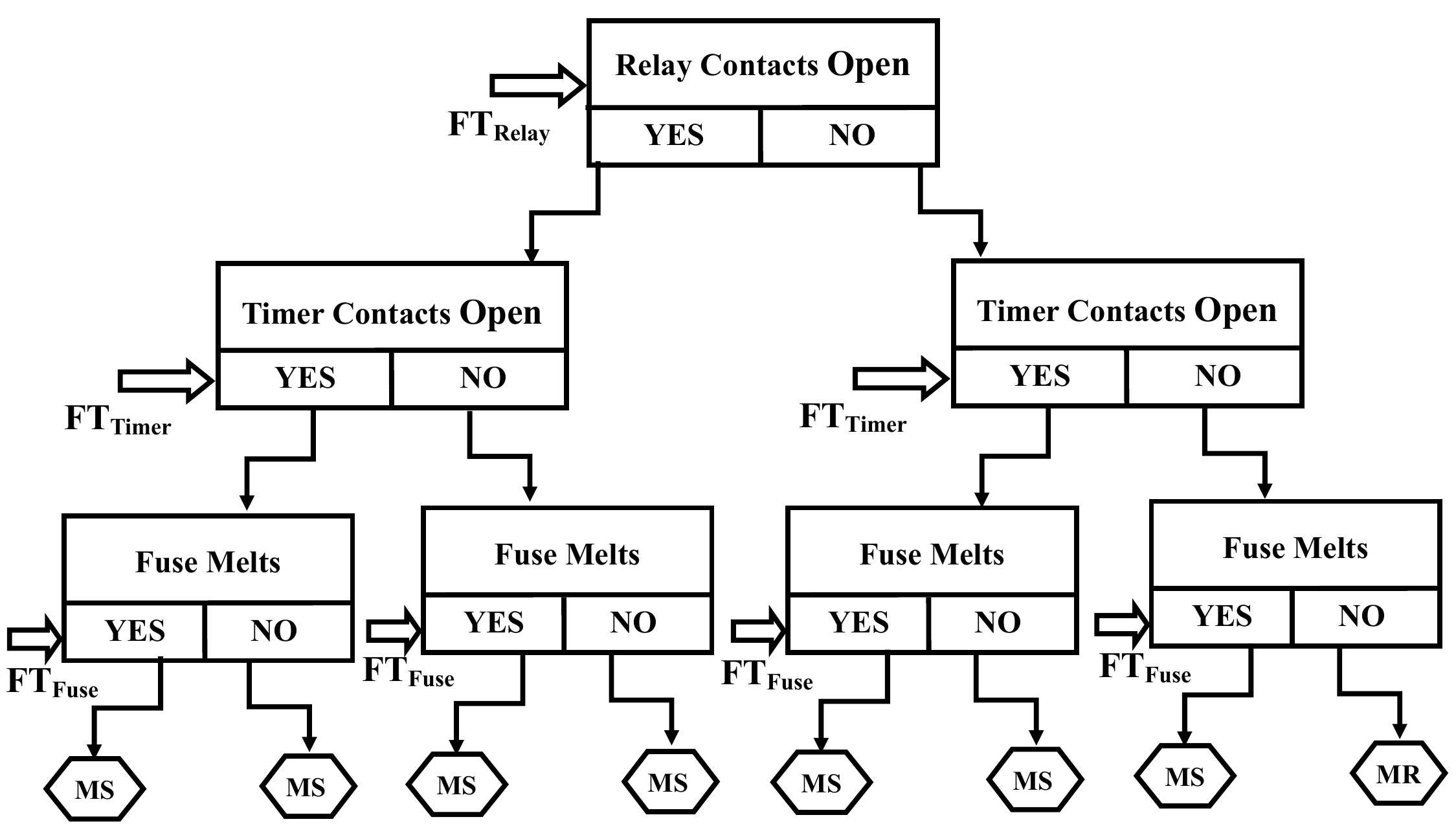}
  \caption{\protect  Complete CCD Model of the MCC}
  \label{Fig: Complete CCD Model}
\end{figure}

\begin{figure}[!h]
  \centering
  \includegraphics[width= 0.39\columnwidth]{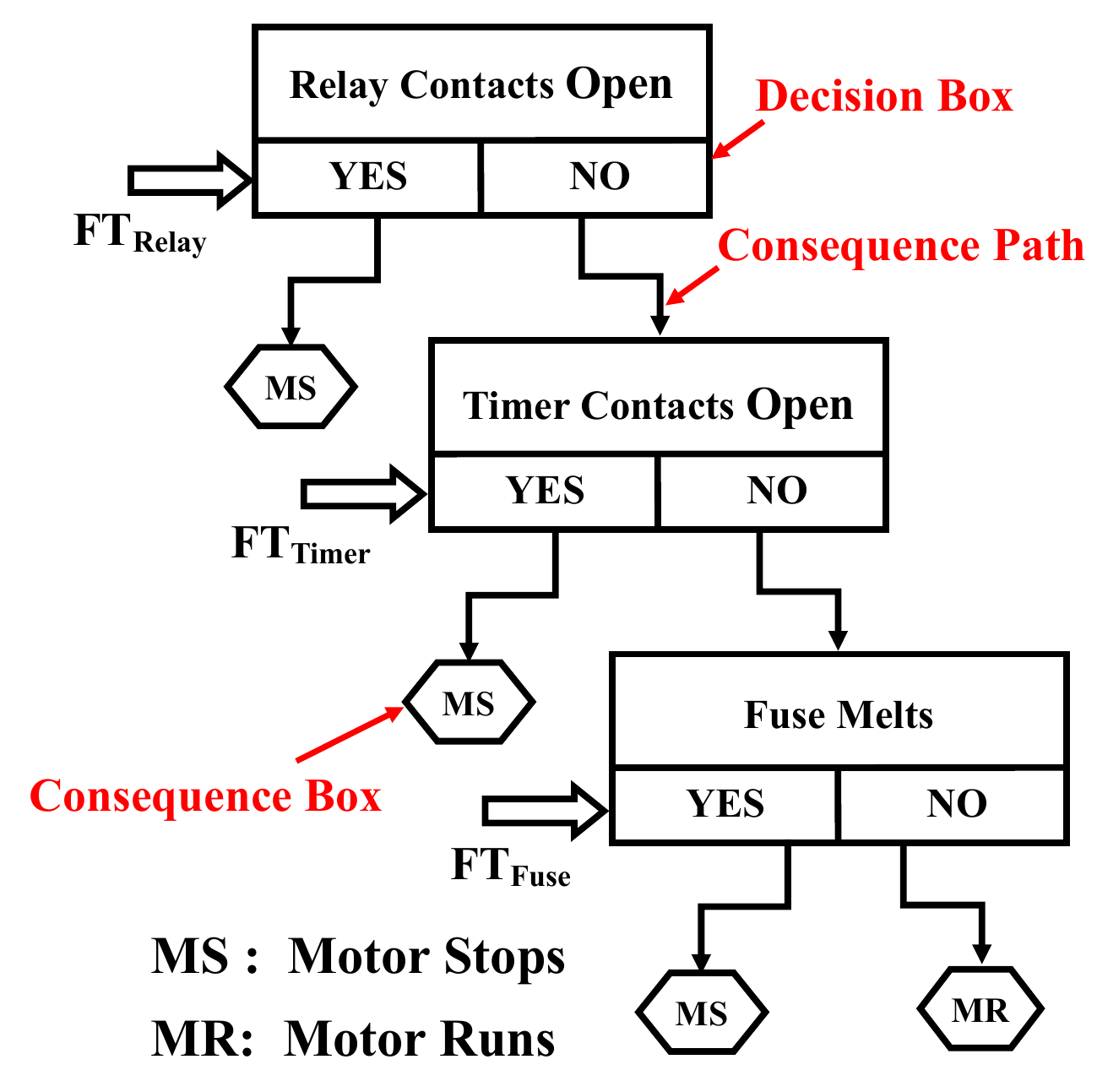}
  \caption{\protect  Reduced CCD Model of the MCC}
  \label{Fig:  Reduced CCD Model}
\end{figure}

\item \textit{CCD probabilistic analysis}: The probabilities of the two consequence boxes MS and MR in Fig. \ref{Fig:  Reduced CCD Model} can be expressed mathematically as:
\begin{equation}
	\label{Eq:probability1} 
	\begin{split}
	\centering 
	\mathcal{P} (Consequence\_Box_{MS})   & = 
	\mathcal{P} (Relay_{S}) + \mathcal{P} (Relay_{F}) \times \mathcal{P} (Timer_{S}) + \\ & \hspace{4mm} \mathcal{P} (Relay_{F}) \times \mathcal{P} (Timer_{F}) \times \mathcal{P} (Fuse_{S}) 
   \end{split}
\end{equation}
\begin{equation}
	\label{Eq:probability2}
	\centering 
	\mathcal{P} (Consequence\_Box_{MR}) = 
	\mathcal{P} (Relay_{F}) \times  \mathcal{P} (Timer_{F}) \times  \mathcal{P} (Fuse_{F}) 
\end{equation}

\noindent where $\mathcal{P} (\mathcal{X}_F)$ is the unreliability function or the probability of failure for a component~$\mathcal{X}$, i.e., $\mathrm{FT}_\mathcal{X}$ model, and $\mathcal{P} (\mathcal{X}_S)$ is the reliability function or the probability of operating, i.e., the complement of the $\mathrm{FT}_\mathcal{X}$ model.\\
\end{enumerate}

In the next section, we present, in detail, the formalization of CCDs in the HOL4 theorem prover to analyze the failures at the subsystem level of a given safety-critical complex system and determine all their possible cascading dependencies of complete/partial reliability and failure events that can occur at the system~level.


\subsection{Formal CCD Modeling}

We start the formalization of CDDs by formally model its basic symbols, as described in~Table~\ref{Table: CCD Symbol} in HOL4 as follows:
\begin{flushleft}
\label{Definition 3}
	\texttt{\bf{Definition 3:}}\\
	\vspace{1pt} 
	\small{\texttt{$\vdash$ DEC\_BOX p X Y =  if X = 1 then FST Y else if X = 0 then SND Y 
	else p\_space p}}    	
\end{flushleft}

\noindent where \texttt{Y} is an ordered pair (\texttt{FST Y}, \texttt{SND Y}) representing the reliability and unreliability functions in a decision box, respectively. The condition \texttt{X = 1} represents the \texttt{YES Box} while \texttt{X = 0} represents the \texttt{NO Box}. If \texttt{X} is neither \texttt{1} nor \texttt{0}, for instance,  \texttt{X = 2}, this represents the irrelevance of the decision box, which returns the probability space \textit{p} to be used in the reduction process of~CCDs.\\

Secondly, we define the CCD \textit{Consequence path} by recursively applying the \texttt{BRANCH} ET constructor on a given $\mathcal{N}$~list of decision boxes ($\mathrm{\small{\texttt{DEC\_BOX}}}_\mathcal{N}$) using the HOL4 recursive function \texttt{FOLDL} as: 
\begin{flushleft}
\label{Definition 4}
	\texttt{\bf{Definition 4:}}\\
	\vspace{1pt} 
	\small{\texttt{$\vdash$ CONSEQ\_PATH p ($\mathrm{	\small{\texttt{DEC\_BOX}}}_1$::$\mathrm{\small{\texttt{DEC\_BOX}}}_\mathcal{N}$)  = \\ \qquad FOLDL ($\lambda$a b. ETREE (BRANCH a b))  $\mathrm{\small{\texttt{DEC\_BOX}}}_1$ $\mathrm{\small{\texttt{DEC\_BOX}}}_\mathcal{N}$}}    	
\end{flushleft}

Finally, we define the CCD \textit{Consequence box} by mapping the function \texttt{CONSEQ\_PATH} on a list using the HOL4 function \texttt{MAP}, then applies the \texttt{NODE} ET constructor:   
\begin{flushleft}
\label{Definition 5}
	\texttt{\bf{Definition 5:}}\\
	\vspace{1pt} 
	\small{\texttt{$\vdash$ CONSEQ\_BOX p $\mathrm{L}_\mathcal{M}$ =  ETREE (NODE (MAP ($\lambda$a. CONSEQ\_PATH p a) $\mathrm{L}_\mathcal{M}$))}}    	
\end{flushleft}

Using the above definitions, we can construct a complete CCD model (\textit{Step~2} in Fig. \ref{Fig: CCD}) for the MCC system shown in Fig.~\ref{Fig: Complete CCD Model}, in HOL4 as: 

\begin{flushleft}
	\vspace{1pt} 
	\small{\texttt{$\vdash$ MCC\_COMPLETE\_CCD $\mathrm{FT}_{R}$ $\mathrm{FT}_{T}$ $\mathrm{FT}_{F}$ = \\ \quad 
	CONSEQ\_BOX p \\ \quad [[DEC\_BOX p 1 ($\overline{\mathrm{FT}_{R}}$,$\mathrm{FT}_{R}$);DEC\_BOX p 1 ($\overline{\mathrm{FT}_{T}}$,$\mathrm{FT}_{T}$);DEC\_BOX p 1 ($\overline{\mathrm{FT}_{F}}$,$\mathrm{FT}_{F}$)]; \\ \quad  \hspace{0.01mm} [DEC\_BOX p 1 ($\overline{\mathrm{FT}_{R}}$,$\mathrm{FT}_{R}$);DEC\_BOX p 1 ($\overline{\mathrm{FT}_{T}}$,$\mathrm{FT}_{T}$);DEC\_BOX p 0 ($\overline{\mathrm{FT}_{F}}$,$\mathrm{FT}_{F}$)]; \\ \quad  \hspace{0.01mm}  [DEC\_BOX p 1 ($\overline{\mathrm{FT}_{R}}$,$\mathrm{FT}_{R}$);DEC\_BOX p 0 ($\overline{\mathrm{FT}_{T}}$,$\mathrm{FT}_{T}$);DEC\_BOX p 1 ($\overline{\mathrm{FT}_{F}}$,$\mathrm{FT}_{F}$)];  \\ \quad  \hspace{0.01mm}  [DEC\_BOX p 0 ($\overline{\mathrm{FT}_{R}}$,$\mathrm{FT}_{R}$);DEC\_BOX p 1 ($\overline{\mathrm{FT}_{T}}$,$\mathrm{FT}_{T}$);DEC\_BOX p 0 ($\overline{\mathrm{FT}_{F}}$,$\mathrm{FT}_{F}$)]; \\ \quad  \hspace{0.01mm} [DEC\_BOX p 0 ($\overline{\mathrm{FT}_{R}}$,$\mathrm{FT}_{R}$);DEC\_BOX p 0 ($\overline{\mathrm{FT}_{T}}$,$\mathrm{FT}_{T}$);DEC\_BOX p 1 ($\overline{\mathrm{FT}_{F}}$,$\mathrm{FT}_{F}$)]; \\ \quad  \hspace{0.01mm} 	[DEC\_BOX p 0 ($\overline{\mathrm{FT}_{R}}$,$\mathrm{FT}_{R}$);DEC\_BOX p 0 ($\overline{\mathrm{FT}_{T}}$,$\mathrm{FT}_{T}$);DEC\_BOX p 0 ($\overline{\mathrm{FT}_{F}}$,$\mathrm{FT}_{F}$)]]}}
\end{flushleft}

In CCD analysis \cite{vyzaite2006cause}, \textit{Step~3} in Fig. \ref{Fig: CCD} is used to model the accurate functional behavior of systems in the sense that the irrelevant decision boxes should be removed from a complete CCD model. Upon this, the actual CCD model of the MCC system after reduction, as shown in Fig.~\ref{Fig:  Reduced CCD Model}, can be obtained by assigning \texttt{X} with neither~\texttt{1} nor \texttt{0}, for instance,  \texttt{X = 2}, which represents the irrelevance of the decision~box,~in HOL4~as:

\begin{flushleft}
	\vspace{1pt} 
	\small{\texttt{$\vdash$ MCC\_REDUCED\_CCD $\mathrm{FT}_{R}$ $\mathrm{FT}_{T}$ $\mathrm{FT}_{F}$ = \\ \quad 
	CONSEQ\_BOX p \\ \quad  [[DEC\_BOX p 1 ($\overline{\mathrm{FT}_{R}}$,$\mathrm{FT}_{R}$);DEC\_BOX p 2 ($\overline{\mathrm{FT}_{T}}$,$\mathrm{FT}_{T}$);DEC\_BOX p 2 ($\overline{\mathrm{FT}_{F}}$,$\mathrm{FT}_{F}$)]; \\ \quad  \hspace{0.01mm}  [DEC\_BOX p 0 ($\overline{\mathrm{FT}_{R}}$,$\mathrm{FT}_{R}$);DEC\_BOX p 1 ($\overline{\mathrm{FT}_{T}}$,$\mathrm{FT}_{T}$);DEC\_BOX p 2 ($\overline{\mathrm{FT}_{F}}$,$\mathrm{FT}_{F}$)]; \\ \quad  \hspace{0.01mm} 	[DEC\_BOX p 0 ($\overline{\mathrm{FT}_{R}}$,$\mathrm{FT}_{R}$);DEC\_BOX p 0 ($\overline{\mathrm{FT}_{T}}$,$\mathrm{FT}_{T}$);DEC\_BOX p 1 ($\overline{\mathrm{FT}_{F}}$,$\mathrm{FT}_{F}$)]; \\ \quad  \hspace{0.01mm} [DEC\_BOX p 0 ($\overline{\mathrm{FT}_{R}}$,$\mathrm{FT}_{R}$);DEC\_BOX p 0 ($\overline{\mathrm{FT}_{T}}$,$\mathrm{FT}_{T}$);DEC\_BOX p 0 ($\overline{\mathrm{FT}_{F}}$,$\mathrm{FT}_{F}$)]]}}
\end{flushleft}

Also, we can formally \textit{verify} the above reduced CCD model of the MCC system, in HOL4 as: 
\begin{flushleft}
	\vspace{1pt} 
	\small{\texttt{$\vdash$ MCC\_REDUCED\_CCD $\mathrm{FT}_{R}$ $\mathrm{FT}_{T}$ $\mathrm{FT}_{F}$ = \\ \quad 
	CONSEQ\_BOX p \\ \quad  [[DEC\_BOX p 1 ($\overline{\mathrm{FT}_{R}}$,$\mathrm{FT}_{R}$)]; \\ \quad  \hspace{0.01mm}  [DEC\_BOX p 0 ($\overline{\mathrm{FT}_{R}}$,$\mathrm{FT}_{R}$);DEC\_BOX p 1 ($\overline{\mathrm{FT}_{T}}$,$\mathrm{FT}_{T}$)]; \\ \quad  \hspace{0.01mm} 	[DEC\_BOX p 0 ($\overline{\mathrm{FT}_{R}}$,$\mathrm{FT}_{R}$);DEC\_BOX p 0 ($\overline{\mathrm{FT}_{T}}$,$\mathrm{FT}_{T}$);DEC\_BOX p 1 ($\overline{\mathrm{FT}_{F}}$,$\mathrm{FT}_{F}$)]; \\ \quad  \hspace{0.01mm} [DEC\_BOX p 0 ($\overline{\mathrm{FT}_{R}}$,$\mathrm{FT}_{R}$);DEC\_BOX p 0 ($\overline{\mathrm{FT}_{T}}$,$\mathrm{FT}_{T}$);DEC\_BOX p 0 ($\overline{\mathrm{FT}_{F}}$,$\mathrm{FT}_{F}$)]]}}
\end{flushleft}
where $\overline{\mathrm{FT}_{X}}$ for a component \textit{X} is the complement of $\mathrm{FT}_{X}$.

\subsection{Formal CCD Analysis}

The important step in the CCD analysis is to determine the probability of each consequence path occurrence in the CCD~\cite{andrews2002application}. For that purpose, we formally verify the following CCD \textit{generic} probabilistic properties, in HOL4~as follows: \\ 

\begin{figure}[!b]
	\includegraphics[width= 1 \columnwidth]{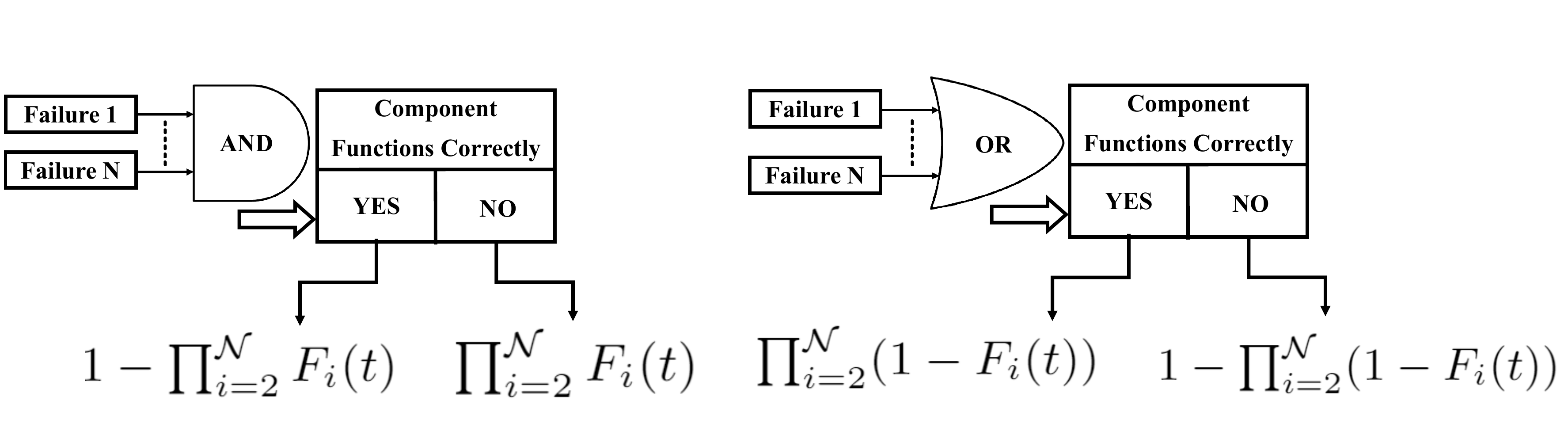} 
	\centering
	\caption{\protect   Decision Boxes with FT Gates}
	\label{Fig: Decision Boxes with FT gates}
\end{figure}

\textit{Property 1}: The probability of a consequence path for \textit{one} decision box assigned with a \textit{generic} FT model, i.e., OR or AND, as shown in Fig. \ref{Fig: Decision Boxes with FT gates}, under the assumptions described in Table \ref{Table: Probability of Failures of Fault Tree Gates}, respectively as follows:
\begin{flushleft}
\label{Theorem 1}
	\texttt{\bf{Theorem 1:}}\\
	\vspace{1pt} 
	\small{\texttt{$\vdash$ \texttt{prob\_space~p} $\wedge$ \texttt{$F_\mathcal{N}$ $\in$ events p} $\wedge$ \texttt{MUTUAL\_INDEP p $F_\mathcal{N}$} $\Rightarrow$  \\  \hspace{3mm}
prob p  \\ \qquad (CONSEQ\_PATH p  [DEC\_BOX p X (FTree p (NOT (OR $\mathrm{F}_\mathcal{N}$)),FTree p (OR $\mathrm{F}_\mathcal{N}$))]) =  \\  \qquad \hspace{0.1mm}   if X = 1 then  $\prod$ (PROB\_LIST p (COMPL\_LIST p $\mathrm{F}_\mathcal{N}$))      \\  else if X = 0 then   1 - $\prod$ (PROB\_LIST p (COMPL\_LIST p $\mathrm{F}_\mathcal{N}$))
 else  1}}
\end{flushleft}

For example, consider a system \texttt{X} consists of two components $\mathrm{C}_{1}$ and $\mathrm{C}_{2}$. Assuming the failure of either one  them causes the system failure, i.e.,~$\mathrm{C}_{1F}$ or $\mathrm{C}_{2F}$,  We can formally model the FT of the system ($\mathrm{FT}_{system}$), in HOL4~as:
\begin{flushleft}
  \vspace{1pt} \small{\texttt{$\vdash$ 
     $\mathrm{FT}_{system}$ p $\mathrm{C}_{1F}$ $\mathrm{C}_{2F}$ = FTree p (OR [$\mathrm{C}_{1F}$;$\mathrm{C}_{2F}$])}}
\end{flushleft}

\noindent Using Theorem 1, we can obtain the probability of a decision box \texttt{YES/NO} outcomes connected to the above FT model, respectively, in HOL4 as: 
\begin{flushleft}
	\vspace{1pt} 
	\small{\texttt{$\vdash$
prob p (CONSEQ\_PATH p [DEC\_BOX p 1 ($\overline{\mathrm{FT}_{system}}$,$\mathrm{FT}_{system}$))]) = \\  \quad  \hspace{2mm} (1 - prob p $\mathrm{C}_{1F}$) $\times$ (1 -  prob p $\mathrm{C}_{2F}$)}}
\end{flushleft}

\begin{flushleft}
	\vspace{1pt} 
	\small{\texttt{$\vdash$
prob p (CONSEQ\_PATH p [DEC\_BOX p 0 ($\overline{\mathrm{FT}_{system}}$,$\mathrm{FT}_{system}$))]) = \\  \quad  \hspace{2mm} 1 - (1 - prob p $\mathrm{C}_{1F}$) $\times$ (1 -  prob p $\mathrm{C}_{2F}$)}}
\end{flushleft}

\begin{flushleft}
\label{Theorem 2}
	\texttt{\bf{Theorem 2:}}\\
	\vspace{1pt} 
	\small{\texttt{$\vdash$ \texttt{prob\_space~p} $\wedge$ \texttt{$F_\mathcal{N}$ $\in$ events p} $\wedge$ \texttt{MUTUAL\_INDEP p $F_\mathcal{N}$} $\Rightarrow$  \\  \hspace{3mm}
prob p  \\ \qquad  (CONSEQ\_PATH p \\ \qquad  \qquad \qquad  [DEC\_BOX p X (FTree p (NOT (AND $\mathrm{F}_\mathcal{N}$)),FTree p (AND $\mathrm{F}_\mathcal{N}$))]) = \\  \qquad \hspace{4.2mm} if X = 1 then 1 - $\prod$ (PROB\_LIST p $\mathrm{F}_\mathcal{N}$)
     \\ \quad  else if X = 0 then $\prod$ (PROB\_LIST p $\mathrm{F}_\mathcal{N}$) else 1}}
\end{flushleft}

For instance, in the above example, assume the failure of both components simultaneously only causes the system failure, i.e.,~$\mathrm{C}_{1F}$ and $\mathrm{C}_{2F}$. We can formally model the FT of the system, in HOL4~as:
\begin{flushleft}
  \vspace{1pt} \small{\texttt{$\vdash$ 
     $\mathrm{FT}_{system}$ p $\mathrm{C}_{1F}$ $\mathrm{C}_{2F}$ = FTree p (AND[$\mathrm{C}_{1F}$;$\mathrm{C}_{2F}$])}}
\end{flushleft}

\begin{figure}[!b]
	\includegraphics[width= 1 \columnwidth]{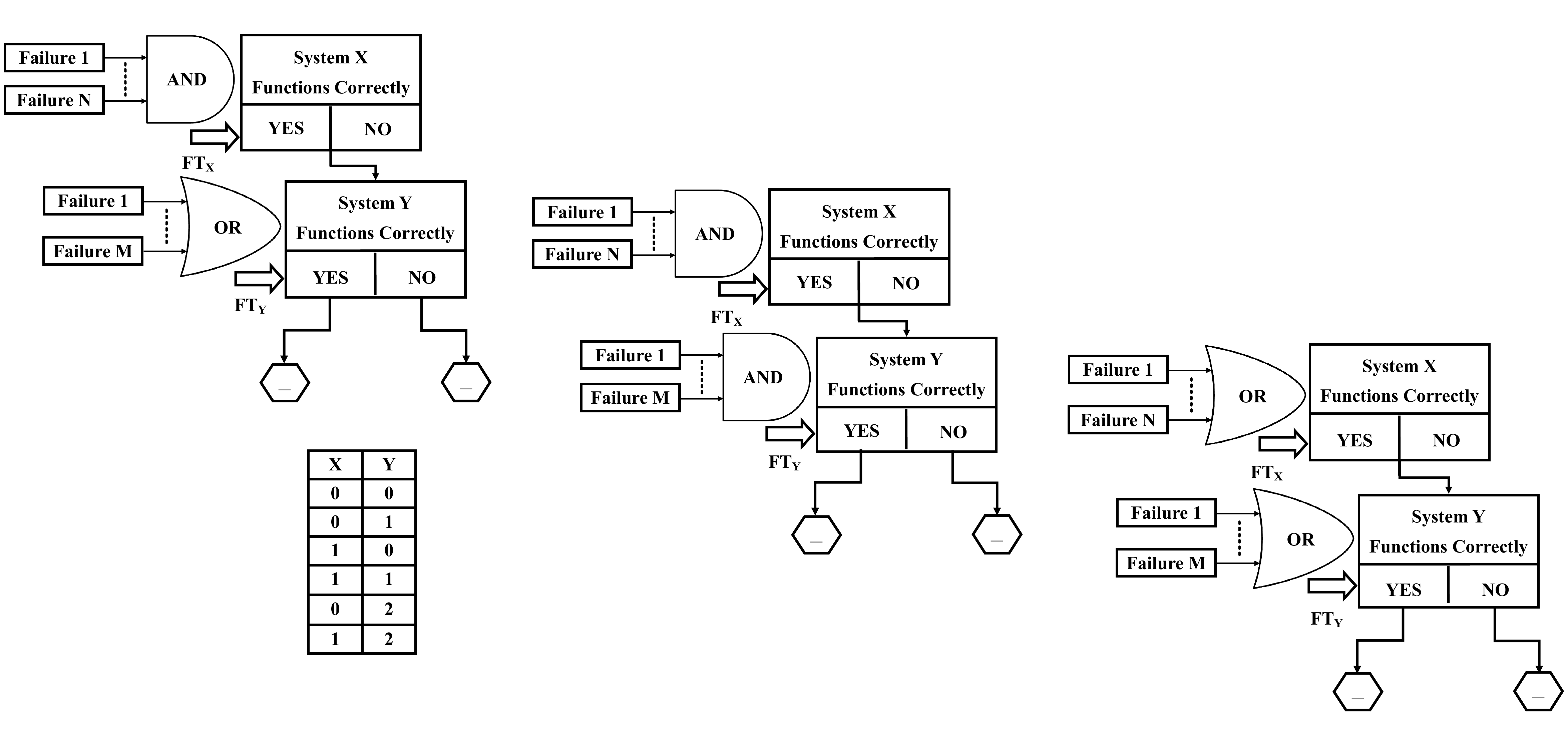} 
	\centering
	\caption{\protect   Two-level Decision Boxes for CCD Analysis}
	\label{Fig: Two level of DECISION BOX}
\end{figure}

\noindent Using Theorem 2, we can obtain the probability of a decision box \texttt{YES/NO} outcomes connected to the above FT model, respectively, in HOL4 as: 
\begin{flushleft}
	\vspace{1pt} 
	\small{\texttt{$\vdash$
prob p (CONSEQ\_PATH p [DEC\_BOX p 1 ($\overline{\mathrm{FT}_{system}}$,$\mathrm{FT}_{system}$))]) = \\  \quad  \hspace{3mm} 1 - prob p $\mathrm{C}_{1F}$ $\times$ prob p $\mathrm{C}_{2F}$}}
\end{flushleft}

\begin{flushleft}
	\vspace{1pt} 
	\small{\texttt{$\vdash$
prob p (CONSEQ\_PATH p [DEC\_BOX p 0 ($\overline{\mathrm{FT}_{system}}$,$\mathrm{FT}_{system}$))]) = \\  \quad  \hspace{3mm} prob p $\mathrm{C}_{1F}$ $\times$ prob p $\mathrm{C}_{2F}$}}
\end{flushleft}

\textit{Property 2}: The probability of \textit{two}-level decision boxes assigned to a CCD path with all combinations of FT gates (\texttt{AND}-\texttt{OR}/\texttt{OR}-\texttt{AND} , \texttt{AND}-\texttt{AND} and \texttt{OR}-\texttt{OR}), as shown in Fig. \ref{Fig: Two level of DECISION BOX}. Each combination has 4 possible operating scenarios that can occur (\texttt{0-0}, \texttt{0-1}, \texttt{1-0} and \texttt{1-1}) and 2 other possible reduction scenarios that may be required in \textit{Step}~3 (\texttt{0-2} and \texttt{1-2}), which represents the removal of the decision box~Y from the path. The basic idea is to select different combinations of  decision boxes to achieve the desired system behavior and also select the reduction combination ($>$ 1) to remove irreverent decision boxes. This probabilistic expressions can be formally verified, in HOL4 as:
\vspace{-2mm}

\begin{flushleft}
\label{Theorem 3}
	\texttt{\bf{Theorem 3:}}\\
	\vspace{1pt} 
	\small{\texttt{$\vdash$ 
	 prob\_space p $\wedge$ ($\forall$y. y $\in$ ($\mathrm{F}_\mathcal{N}$++$\mathrm{F}_\mathcal{M}$) $\Rightarrow$ y $\in$ events p)  $\wedge$ \\  \hspace{2mm} 
	 MUTUAL\_INDEP p   ($\mathrm{F}_\mathcal{N}$++$\mathrm{F}_\mathcal{M}$)
	  $\Rightarrow$  \\  \hspace{3mm}
	  prob p (CONSEQ\_PATH p \\  \qquad  \qquad 
          [DEC\_BOX p X (FTree p (NOT (AND $\mathrm{F}_\mathcal{N}$)),FTree p (AND $\mathrm{F}_\mathcal{N}$));  \\   \qquad  \qquad   \hspace{0.3mm}
  	   DEC\_BOX p Y (FTree p (NOT (OR $\mathrm{F}_\mathcal{M}$)),FTree p (OR $\mathrm{F}_\mathcal{M}$))]) =  \\ \quad \quad  \hspace{3.5mm}
   if X = 0  $\wedge$  Y = 0 then \\ \quad \quad  $\prod$ (PROB\_LIST p $\mathrm{F}_\mathcal{N}$) $\times$  (1 - $\prod$ (PROB\_LIST p (COMPL\_LIST p $\mathrm{F}_\mathcal{M}$))) \\ \quad else if X = 0  $\wedge$  Y = 1 then \\ \quad \quad  $\prod$ (PROB\_LIST p $\mathrm{F}_\mathcal{N}$) $\times$   $\prod$ (PROB\_LIST p (COMPL\_LIST p $\mathrm{F}_\mathcal{M}$)) \\ \quad
   else if X = 1  $\wedge$  Y = 0 then  \\ \quad \hspace{0.5mm} (1 - $\prod$ (PROB\_LIST p $\mathrm{F}_\mathcal{N}$)) $\times$ (1 - $\prod$ (PROB\_LIST p (COMPL\_LIST p $\mathrm{F}_\mathcal{M}$))) \\ \quad 
   else if X = 1  $\wedge$  Y = 1 then  \\ \quad \hspace{0.5mm} (1 - $\prod$ (PROB\_LIST p $\mathrm{F}_\mathcal{N}$)) $\times$ $\prod$ (PROB\_LIST p (COMPL\_LIST p $\mathrm{F}_\mathcal{M}$)) \\ \quad
   else if X = 0  $\wedge$  Y = 2 then  $\prod$ (PROB\_LIST p $\mathrm{F}_\mathcal{N}$)  \\ \quad 
    else if X = 1  $\wedge$  Y = 2 then   (1 - $\prod$ (PROB\_LIST p $\mathrm{F}_\mathcal{N}$)) else 1}}
\end{flushleft}

\vspace{-6mm}

\begin{flushleft}
\label{Theorem 4}
	\texttt{\bf{Theorem 4:}}\\
	\vspace{1pt} 
	\small{\texttt{$\vdash$ 
	  prob p (CONSEQ\_PATH p \\ \qquad \qquad 
          [DEC\_BOX p X (FTree p (NOT (AND $\mathrm{F}_\mathcal{N}$)),FTree p (AND $\mathrm{F}_\mathcal{N}$)); \\ \qquad \qquad \hspace{0.3mm}
  	   DEC\_BOX p Y (FTree p (NOT (AND $\mathrm{F}_\mathcal{M}$)),FTree p (AND $\mathrm{F}_\mathcal{M}$))]) =  \\ \quad  \quad \hspace{4.2mm} 
   if X = 0  $\wedge$  Y = 0 then  \\ \quad \quad $\prod$ (PROB\_LIST p $\mathrm{F}_\mathcal{N}$) $\times$ $\prod$ (PROB\_LIST p $\mathrm{F}_\mathcal{M}$)   \\ \quad 
    else if X = 0  $\wedge$  Y = 1 then   \\ \quad \quad $\prod$ (PROB\_LIST p $\mathrm{F}_\mathcal{N}$) $\times$ (1 - $\prod$ (PROB\_LIST p $\mathrm{F}_\mathcal{M}$)) 
   \\ \qquad \qquad  \qquad  \qquad  \qquad \qquad  \vdots 
    \\ \quad 
    else if X = 1  $\wedge$  Y = 2 then  (1 - $\prod$ (PROB\_LIST p $\mathrm{F}_\mathcal{N}$))  else  1}}
\end{flushleft}

\vspace{-6mm}

\begin{flushleft}
\label{Theorem 5}
	\texttt{\bf{Theorem 5:}}\\
	\vspace{1pt} 
	\small{\texttt{$\vdash$ 
	  prob p (CONSEQ\_PATH p  \\ \qquad \qquad 
          [DEC\_BOX p X (FTree p (NOT (OR $\mathrm{F}_\mathcal{N}$)),FTree p (OR $\mathrm{F}_\mathcal{N}$));  \\ \qquad \qquad \hspace{0.3mm}
  	   DEC\_BOX p Y (FTree p (NOT (OR $\mathrm{F}_\mathcal{M}$)),FTree p (OR $\mathrm{F}_\mathcal{M}$))]) =  \\ \quad \quad  \hspace{4.5mm}
   if X = 0  $\wedge$  Y = 0 then \\ \quad (1 - $\prod$ (PROB\_LIST p (COMPL\_LIST p $\mathrm{F}_\mathcal{N}$))) $\times$ \\ \quad  (1 - $\prod$ (PROB\_LIST p (COMPL\_LIST p $\mathrm{F}_\mathcal{M}$))) 
       \\ \quad else if X = 0  $\wedge$  Y = 1 then  \\ \quad (1 - $\prod$ (PROB\_LIST p (COMPL\_LIST p $\mathrm{F}_\mathcal{N}$))) $\times$ \\ \quad  $\prod$ (PROB\_LIST p (COMPL\_LIST p $\mathrm{F}_\mathcal{M}$))
    \\ \qquad \qquad  \qquad  \qquad  \qquad \qquad   \vdots 
    \\ \quad 
    else if X = 1  $\wedge$  Y = 2 then  $\prod$ (PROB\_LIST p (COMPL\_LIST p $\mathrm{F}_\mathcal{N}$)) else 1}}
\end{flushleft}

\begin{figure}[!t]
	\includegraphics[width= 0.85 \columnwidth]{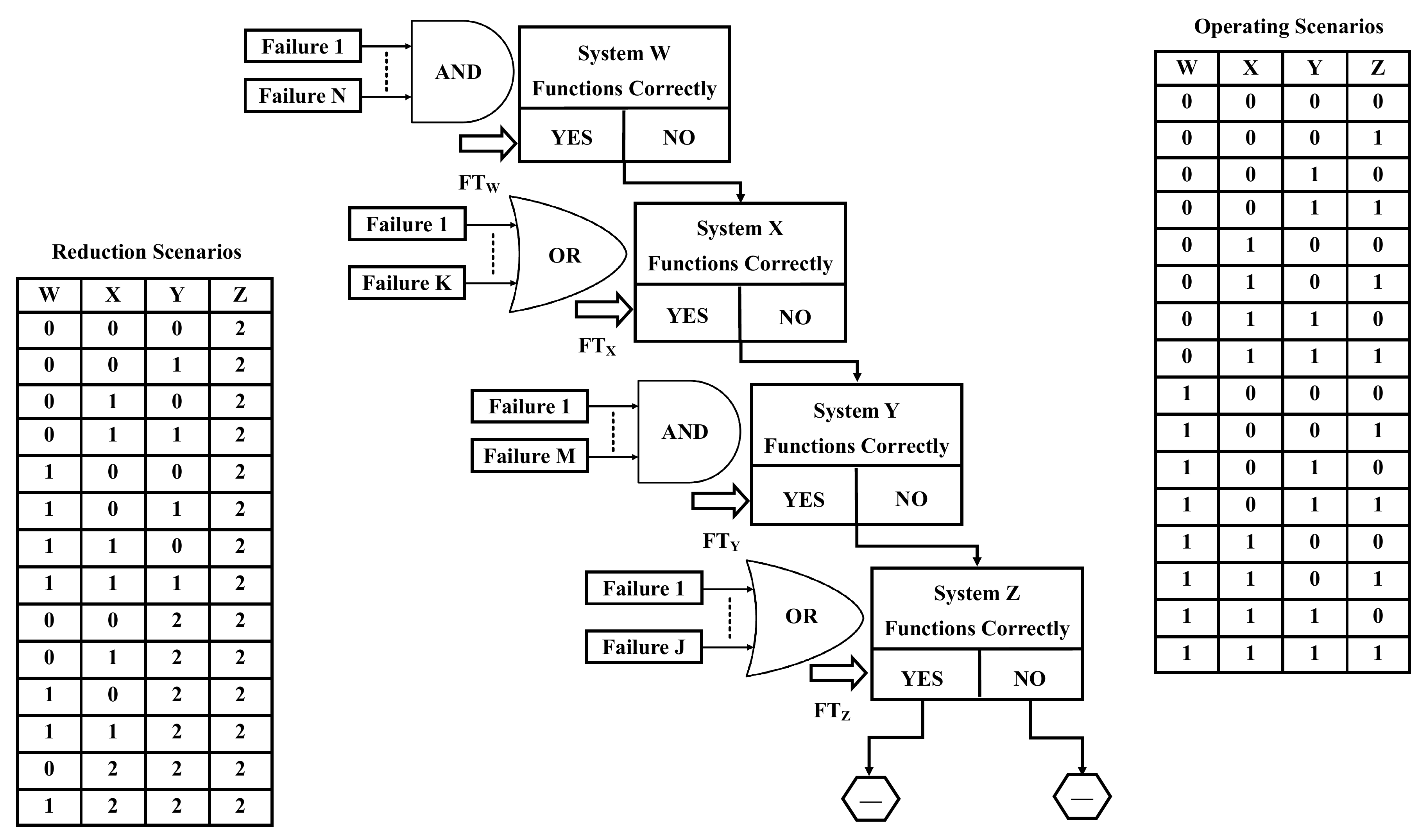} 
	\centering
	\caption{\protect  Four-level Decision Boxes for CCD Analysis}
	\label{Fig:  Four-level Decision Boxes for CCD Analysis}
\end{figure}

\textit{Property 3}:  A \textit{generic} probabilistic property for a consequence path consisting of complex \textit{four}-level decision boxes associated with  different combination of FTs and each one consisting of $\mathcal{N}$ components  (\texttt{AND}-\texttt{OR}-\texttt{AND}-\texttt{OR}/\texttt{OR}-\texttt{AND}-\texttt{OR}-\texttt{AND}/\texttt{AND}-\texttt{AND}-\texttt{OR}-\texttt{OR}/\texttt{OR}-\texttt{OR}-\texttt{AND}-\texttt{AND}), which has 16 possible operating scenarios that can occur and  14 other possible reduction possibilities, as shown in Fig.~\ref{Fig:  Four-level Decision Boxes for CCD Analysis}, in HOL4~as:

\vspace{-2mm}

\begin{flushleft}
\label{Theorem 6}
	\texttt{\bf{Theorem 6:}}\\
	\vspace{1pt} 
	\small{\texttt{$\vdash$ \small{\texttt{Let}} \\ 
	\texttt{$\small{\texttt{W}}_\mathrm{F}$ = $\prod$ (PROB\_LIST p $\mathrm{F}_\mathcal{N}$); \\  $\overline{\small{\texttt{W}}}$ \hspace{-0.3mm} =  1 - $\small{\texttt{W}}_\mathrm{F}$}}; \\
	\texttt{$\small{\texttt{X}}_\mathrm{F}$ = 1 - $\prod$ (PROB\_LIST p (COMPL\_LIST p $\mathrm{F}_\mathcal{K}$)); $\overline{\small{\texttt{X}}}$ \hspace{-0.3mm} = 1 - $\small{\texttt{X}}_\mathrm{F}$};\\  
	\texttt{$\small{\texttt{Y}}_\mathrm{F}$ = $\prod$ (PROB\_LIST p $\mathrm{F}_\mathcal{M}$); \\ 
	$\overline{\small{\texttt{Y}}}$  \hspace{-0.3mm} = 1 - $\small{\texttt{Y}}_\mathrm{F}$};\\ 
	\texttt{$\small{\texttt{Z}}_\mathrm{F}$ = 1 - $\prod$ (PROB\_LIST p (COMPL\_LIST p $\mathrm{F}_\mathcal{J}$)); $\overline{\small{\texttt{Z}}}$ \hspace{-0.3mm} = 1 - $\small{\texttt{Z}}_\mathrm{F}$}} \\
	\small{\texttt{in \\ \quad 
	\vspace{1pt} 	  
	  prob p \\ \qquad (CONSEQ\_PATH p \\ \qquad \qquad 
          [DEC\_BOX p W (FTree p (NOT (AND $\mathrm{F}_\mathcal{N}$)),FTree p (AND $\mathrm{F}_\mathcal{N}$)); \\ \qquad \qquad  \hspace{0.3mm}
          DEC\_BOX p X (FTree p (NOT (OR $\mathrm{F}_\mathcal{K}$)),FTree p (OR $\mathrm{F}_\mathcal{K}$)); \\  \qquad \qquad  \hspace{0.3mm}
          DEC\_BOX p Y (FTree p (NOT (AND $\mathrm{F}_\mathcal{M}$)),FTree p (AND $\mathrm{F}_\mathcal{M}$)); \\  \qquad \qquad  \hspace{0.3mm}
  	      DEC\_BOX p Z (FTree p (NOT (OR $\mathrm{F}_\mathcal{J}$)),FTree p (OR $\mathrm{F}_\mathcal{J}$))]) =  \\ \quad \quad  \hspace{4mm}
   if W = 0  $\wedge$  X = 0  $\wedge$  Y = 0  $\wedge$  Z = 0  \\ \quad then
   $\small{\texttt{W}}_\mathrm{F}$ $\times$ $\small{\texttt{X}}_\mathrm{F}$ $\times$ $\small{\texttt{Y}}_\mathrm{F}$ $\times$ $\small{\texttt{Z}}_\mathrm{F}$ \\ \quad 
   else if W = 0  $\wedge$  X = 0  $\wedge$  Y = 0  $\wedge$  Z = 1 \\ \quad then   
   $\small{\texttt{W}}_\mathrm{F}$ $\times$ $\small{\texttt{X}}_\mathrm{F}$ $\times$ $\small{\texttt{Y}}_\mathrm{F}$ $\times$ $\overline{\small{\texttt{Z}}}$ \\ \quad 
   else if W = 0  $\wedge$  X = 0  $\wedge$  Y = 1  $\wedge$  Z = 0 \\ \quad then   
   $\small{\texttt{W}}_\mathrm{F}$ $\times$ $\small{\texttt{X}}_\mathrm{F}$ $\times$ $\overline{\small{\texttt{Y}}}$ $\times$ $\small{\texttt{Z}}_\mathrm{F}$ 
    \\ \qquad \qquad  \qquad  \qquad  \quad \qquad \qquad  \vdots 
    \\ \quad 
    else if W = 1  $\wedge$  X = 1  $\wedge$  Y = 2  $\wedge$  Z = 2 \\ \quad then
    $\overline{\small{\texttt{W}}}$  $\times$  $\overline{\small{\texttt{X}}}$ 
    \\ \quad
    else if W = 1  $\wedge$  X = 2  $\wedge$  Y = 2  $\wedge$  Z = 2 \\ \quad then $\overline{\small{\texttt{W}}}$ 
   else 1}}
\end{flushleft}

\begin{figure}[!h]
	\includegraphics[width= 0.9 \columnwidth]{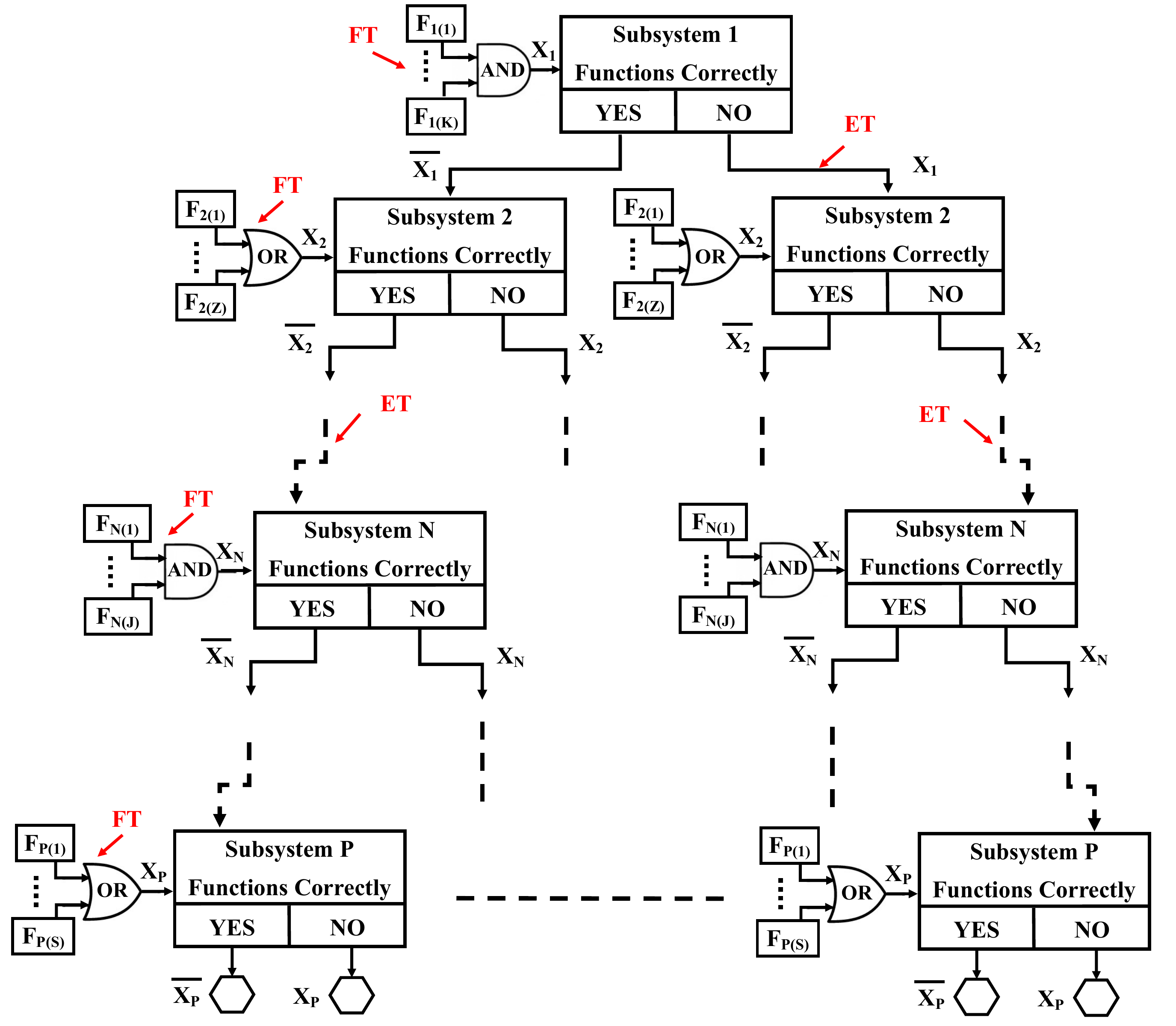} 
	\centering
	\caption{\protect  Generic \textit{$\mathcal{N}$}-level CCD Analysis}
	\label{Fig:  Generic CCD Analysis}
\end{figure}

For  complex  systems  consisting  of \textit{$\mathcal{N}$}-level decision boxes, where each decision box is associated with an AND/OR gate consisting of an arbitrary list of failure events, we define \textit{three} types  \textit{A},  \textit{B} and  \textit{C} of possible CCD outcomes, as shown in Fig. \ref{Fig:  Generic CCD Analysis}, with a new proposed mathematics as:\\

\textit{Property 4 (N Decision Boxes of Type A):} The probability of $n$ decision boxes assigned to a consequence path corresponding to  $n$ subsystems, where all decision boxes are associated with FT AND models consisting of arbitrary lists of $k$ events, can be expressed mathematically at a specific time \textit {t} for \textit{three} cases~as: \\

\vspace{-2mm}

\noindent (A1)  All outcomes of $n$ decisions boxes are NO 

\begin{equation}
	\label{Equation 6} 
	\centering 
	\mathcal{F}_{A1} (t) =   \prod\limits^n_{i=1} 
	\prod\limits^k_{j=2} {\mathcal{F}}_{ij} (t)  
\end{equation}

\noindent (A2) All outcomes of  $n$  decisions boxes are YES

\begin{equation}
	\label{Equation 7} 
	\centering 
	\mathcal{F}_{A2}  (t) =   \prod\limits^n_{i=1} (1 - 
	\prod\limits^k_{j=2} {\mathcal{F}}_{ij} (t))  
\end{equation}    
    
\noindent (A3) Some outcomes of $m$ decisions boxes are NO and the rest outcomes of $p$ decisions boxes are YES 
    
\begin{equation}
	\label{Equation 8} 
	\centering 
	\mathcal{F}_{A3} (t) =   \Bigg(\prod\limits^m_{i=1} 
	\prod\limits^k_{j=2} {\mathcal{F}}_{ij} (t)\Bigg)   \times \Bigg(\prod\limits^p_{i=1}  (1 - 
	\prod\limits^k_{j=2}{\mathcal{F}}_{ij} (t))\Bigg)  
\end{equation}

To verify the correctness of the above-proposed  new safety analysis mathematical formulations in the HOL4 theorem prover,  we define two \textit{generic}  CCD  functions $\mathcal{SS}^{YES}_{AND}$ and $\mathcal{SS}^{NO}_{AND}$ that can recursively generate the outcomes YES and NO of the function \texttt{FTree},  identified by \texttt{gate} constructors \texttt{AND} and \texttt{NOT}, for a given arbitrary list of all subsystems failure events (\texttt{SSN}), respectively, in HOL4 as: 

\begin{flushleft}
	\texttt{\bf{Definition 6:}}\\
	\vspace{1pt} 
	\small{\texttt{$\vdash$ $\mathcal{SS}^{YES}_{AND}$ p (SS::SSN) = FTree p (NOT (AND SS1))::$\mathcal{SS}^{YES}_{AND}$ p SSN }}    	
\end{flushleft}

\begin{flushleft}
	\texttt{\bf{Definition 7:}}\\
	\vspace{1pt} 
	\small{\texttt{$\vdash$ $\mathcal{SS}^{NO}_{AND}$ p (SS1::SSN) =  FTree p (AND SS1)::$\mathcal{SS}^{NO}_{AND}$ p SSN }}    	
\end{flushleft}

\noindent Using above defined functions, we can verify three \textit{two-dimensional} and \textit{scalable} probabilistic properties corresponding to the above-mentioned safety equations Eq.~\ref{Equation 6}, Eq.~\ref{Equation 7}, and Eq.~\ref{Equation 8}, respectively, in HOL4 as: 

\begin{flushleft}
	\texttt{\bf{Theorem 7:}}\\
	\vspace{1pt} 
	\small{\texttt{$\vdash$   prob p  (CONSEQ\_PATH p ($\mathcal{SS}^{NO}_{AND}$ p SSN)) = \\ \hspace{1mm} $\prod$ (MAP ($\lambda$ a. $\prod$ (PROB\_LIST p a)) SSN)}}
\end{flushleft}

\begin{flushleft}
	\texttt{\bf{Theorem 8:}}\\
	\vspace{1pt} 
	\small{\texttt{$\vdash$  prob p (CONSEQ\_PATH p ($\mathcal{SS}^{YES}_{AND}$ p SSN)) = \\ \hspace{1mm} $\prod$ (MAP ($\lambda$ b. (1 - $\prod$ (PROB\_LIST p b))) SSN)}}
\end{flushleft}

\begin{flushleft}
	\texttt{\bf{Theorem 9:}}\\
	\vspace{1pt} 
	\small{\texttt{$\vdash$ prob p \\ \qquad \hspace{1mm}  (CONSEQ\_PATH p \\ \qquad \qquad \hspace{1mm} [CONSEQ\_PATH p ($\mathcal{SS}^{NO}_{AND}$ p SSm); \\ \hspace{3mm} \\ \qquad \qquad \hspace{3mm} CONSEQ\_PATH p ($\mathcal{SS}^{YES}_{AND}$ p SSp)])  = \\  $\bigg(\prod$ (MAP ($\lambda$ a. $\prod$ (PROB\_LIST p a)) SSm)$\bigg)$ $\times$ \\ $\bigg(\prod$ (MAP ($\lambda$ b. 1 - $\prod$ (PROB\_LIST p b)) SSp)$\bigg)$}}
\end{flushleft}

\textit{Property 5 (N Decision Boxes  of Type B):}  The probabilistic assessment of  $n$ decision boxes assigned to a CCD consequence path, where all decision boxes are associated with \textit{generic} FT OR models consisting of arbitrary lists of $k$ events, can be expressed mathematically for \textit{three} cases:\\

\noindent (B1) All outcomes of $n$ decisions boxes are NO
\begin{equation}
	\label{Equation 10} 
	\centering
	\mathcal{F}_{B1} (t) =   \prod\limits^n_{i=1} 
    (1 - 	\prod\limits^k_{j=2} (1 - {\mathcal{F}}_{ij} (t))) 
\end{equation}

\noindent (B2) All outcomes of  $n$  decisions boxes are YES 

\begin{equation}
	\label{Equation 11} 
	\centering 
	\mathcal{F}_{B2}  (t) =   \prod\limits^n_{i=1} 
	\prod\limits^k_{j=2} (1 - {\mathcal{F}}_{ij} (t))  
\end{equation}

\noindent (B3) Some outcomes of $m$ decisions boxes are NO and some outcomes of $p$ decisions boxes are YES 
 
\begin{equation}
	\label{Equation 12} 
	\centering 
	\mathcal{F}_{B3} (t) =   \Bigg(\prod\limits^m_{i=1} 
    (1 - 	\prod\limits^k_{j=2} (1 - {\mathcal{F}}_{ij} (t)))\Bigg)  \times \Bigg( \prod\limits^p_{i=1} 
	\prod\limits^k_{j=2} (1 - {\mathcal{F}}_{ij} (t))\Bigg)  
\end{equation}

To verify the correctness of the above-proposed  new CCD mathematical formulas in HOL4,  we define two \textit{generic} functions $\mathcal{SS}^{YES}_{OR}$ and $\mathcal{SS}^{NO}_{OR}$ to recursively generate the outcomes YES and NO of the function \texttt{FTree},  identified by \texttt{gate} constructors \texttt{OR} and \texttt{NOT}, for a given list of subsystems events.  

\begin{flushleft}
	\texttt{\bf{Definition 8:}}\\
	\vspace{1pt} 
	\small{\texttt{$\vdash$ $\mathcal{SS}^{YES}_{OR}$ p (SS::SSN) =   FTree p (NOT (OR SS1))::$\mathcal{SS}^{YES}_{OR}$ p SSN }}    	
\end{flushleft}

\begin{flushleft}
	\texttt{\bf{Definition 9:}}\\
	\vspace{1pt} 
	\small{\texttt{$\vdash$ $\mathcal{SS}^{NO}_{OR}$ p (SS1::SSN) = FTree p (OR SS1)::$\mathcal{SS}^{NO}_{OR}$ p SSN }}    	
\end{flushleft}

Using above defined functions, we can formally \textit{verify} three \textit{scalable} probabilistic properties corresponding to Eq.~\ref{Equation 10}, Eq.~\ref{Equation 11}, and Eq.~\ref{Equation 12}, respectively, in HOL4 as: 

\begin{flushleft}
	\texttt{\bf{Theorem 10:}}\\
	\vspace{1pt} 
	\small{\texttt{$\vdash$   prob p  (CONSEQ\_PATH p ($\mathcal{SS}^{NO}_{OR}$ p SSN)) = \\  \hspace{2mm} 
$\prod$ \\  \hspace{3mm} (MAP \\  \hspace{6mm}  ($\lambda$ a. \\  \hspace{9mm}  (1 -  $\prod$  (PROB\_LIST p (compl\_list p a)))) SSN)}}
\end{flushleft}

\begin{flushleft}
	\texttt{\bf{Theorem 11:}}\\
	\vspace{1pt} 
	\small{\texttt{$\vdash$   prob p (CONSEQ\_PATH p ($\mathcal{SS}^{YES}_{OR}$ p SSN)) = \\  \hspace{2mm} 
$\prod$ \\  \hspace{3mm} (MAP \\  \hspace{6mm} ($\lambda$ b. \\  \hspace{9mm}  $\prod$ (PROB\_LIST p (compl\_list p b))) SSN)}}
\end{flushleft}

\begin{flushleft}
	\texttt{\bf{Theorem 12:}}\\
	\vspace{1pt} 
	\small{\texttt{$\vdash$   prob p \\ \qquad \hspace{1mm}  (CONSEQ\_PATH p \\ \qquad \qquad \hspace{1mm} [CONSEQ\_PATH p ($\mathcal{SS}^{NO}_{OR}$ p SSm); \\ \hspace{3mm} \\ \qquad \qquad \hspace{3mm} CONSEQ\_PATH p ($\mathcal{SS}^{YES}_{OR}$ p SSp)])  = \\  \hspace{3mm} 
$\prod$ \\  \hspace{3mm} (MAP \\  \hspace{6mm}  ($\lambda$ a. \\  \hspace{9mm}  (1 -  $\prod$    (PROB\_LIST p  (compl\_list p a)))) SSm) \\ $\times$ 
$\prod$ \\  \hspace{3mm} (MAP \\  \hspace{6mm} ($\lambda$ b. \\  \hspace{9mm}  $\prod$ (PROB\_LIST p (compl\_list p b))) SSp)}}
\end{flushleft}

\hspace{2mm}

\textit{Property 6 (N Decision Boxes  of Type C):} The probabilistic~assessment of $n$ decision boxes assigned to a consequence path for a very complex system, where some $m$ decision boxes are associated with \textit{generic} FT AND models consisting of $k$-events, while other $p$~decision boxes are associated with \textit{generic}~FT OR models consisting of $z$-events, as shown in Fig.~\ref{Fig:  Generic CCD Analysis}, is proposed to be expressed mathematically for \textit{nine} cases as:\\

\noindent (C1)~All outcomes of $m$ and $p$ decisions boxes are NO. 
\begin{equation}
	\label{Equation 13} 
	\centering
    \mathcal{F}_{C1} (t) =   \Bigg(\prod\limits^m_{i=1} 
	\prod\limits^k_{j=2} {\mathcal{F}}_{ij} (t)\Bigg)   \times \Bigg(\prod\limits^p_{i=1} 
    (1 - 	\prod\limits^z_{j=2} (1 - {\mathcal{F}}_{ij} (t)))\Bigg)
\end{equation}

\begin{flushleft}
	\texttt{\bf{Theorem 13:}}\\
	\vspace{1pt} 
	\small{\texttt{$\vdash$   prob p \\ \qquad \hspace{1mm}  (CONSEQ\_PATH p \\ \qquad \qquad \hspace{1mm} [CONSEQ\_PATH p ($\mathcal{SS}^{NO}_{AND}$ p SSm); \\ \hspace{3mm} \\ \qquad \qquad \hspace{3mm} CONSEQ\_PATH p ($\mathcal{SS}^{NO}_{OR}$ p SSp)])  = \\  \hspace{2.5mm} 
	$\prod$ (MAP ($\lambda$ a. $\prod$ (PROB\_LIST p a)) SSm) \\ 
	$\times$ $\prod$ \\  \hspace{3mm} (MAP \\  \hspace{6mm}  ($\lambda$ b. \\  \hspace{9mm}  (1 -  $\prod$ (PROB\_LIST p (compl\_list p b)))) SSp)}}
\end{flushleft}

\hspace{2mm}

\noindent (C2) All outcomes of $m$ and $p$ decisions boxes are YES.
\begin{equation}
	\label{Equation 14} 
	\centering \mathcal{F}_{C2} (t) = \Bigg(\prod\limits^m_{i=1}  (1 - 
	\prod\limits^k_{j=2}{\mathcal{F}}_{ij} (t))\Bigg) \times \Bigg( \prod\limits^p_{i=1} 
	\prod\limits^z_{j=2} (1 - {\mathcal{F}}_{ij} (t))\Bigg)  
\end{equation}

\hspace{2mm}

\begin{flushleft}
	\texttt{\bf{Theorem 14:}}\\
	\vspace{1pt} 
	\small{\texttt{$\vdash$   prob p \\ \qquad \hspace{1mm}  (CONSEQ\_PATH p \\ \qquad \qquad \hspace{1mm} [CONSEQ\_PATH p ($\mathcal{SS}^{YES}_{AND}$ p SSm); \\ \hspace{3mm} \\ \qquad \qquad \hspace{3mm} CONSEQ\_PATH p ($\mathcal{SS}^{YES}_{OR}$ p SSp)])  = \\  \hspace{2.5mm} 
	$\prod$ (MAP ($\lambda$ a. 1 - $\prod$ (PROB\_LIST p a)) SSm) \\ 
	$\times$ $\prod$ \\  \hspace{3mm} (MAP \\  \hspace{6mm} ($\lambda$ b. \\  \hspace{9mm}  $\prod$ (PROB\_LIST p (compl\_list p b))) SSp)}}
\end{flushleft}

\noindent (C3) All outcomes of $m$ decisions boxes are NO and all outcomes of $p$ decisions boxes are YES. 
\begin{equation}
	\label{Equation 15} 
	\centering \mathcal{F}_{C3} (t) =   \Bigg(\prod\limits^m_{i=1} 
	\prod\limits^k_{j=2} {\mathcal{F}}_{ij} (t)\Bigg)   \times \Bigg( \prod\limits^p_{i=1} 
	\prod\limits^z_{j=2} (1 - {\mathcal{F}}_{ij} (t))\Bigg)  
\end{equation}

\begin{flushleft}
	\texttt{\bf{Theorem 15:}}\\
	\vspace{1pt} 
	\small{\texttt{$\vdash$   prob p \\ \qquad \hspace{1mm}  (CONSEQ\_PATH p \\ \qquad \qquad \hspace{1mm} [CONSEQ\_PATH p ($\mathcal{SS}^{NO}_{AND}$ p SSm); \\ \hspace{3mm} \\ \qquad \qquad \hspace{3mm} CONSEQ\_PATH p ($\mathcal{SS}^{YES}_{OR}$ p SSp)])  = \\  \hspace{2.5mm} 
	$\prod$ (MAP ($\lambda$ a. $\prod$ (PROB\_LIST p a)) SSm) \\ 
	$\times$ $\prod$ \\  \hspace{3mm} (MAP \\  \hspace{6mm} ($\lambda$ b. \\  \hspace{9mm}  $\prod$ (PROB\_LIST p (compl\_list p b))) SSp)}}
\end{flushleft}

\noindent (C4) All outcomes of $m$ decisions boxes are YES and all outcomes of $p$ decisions boxes are NO.
\begin{equation}
	\label{Equation 16} 
	\centering  
	\mathcal{F}_{C4} (t) = \Bigg(\prod\limits^m_{i=1}  (1 - 
	\prod\limits^k_{j=2}{\mathcal{F}}_{ij} (t))\Bigg) \times \Bigg(\prod\limits^p_{i=1} 
    (1 - 	\prod\limits^z_{j=2} (1 - {\mathcal{F}}_{ij} (t)))\Bigg)
\end{equation}

\begin{flushleft}
	\texttt{\bf{Theorem 16:}}\\
	\vspace{1pt} 
	\small{\texttt{$\vdash$   prob p \\ \qquad \hspace{1mm}  (CONSEQ\_PATH p \\ \qquad \qquad \hspace{1mm} [CONSEQ\_PATH p ($\mathcal{SS}^{YES}_{AND}$ p SSm); \\ \hspace{3mm} \\ \qquad \qquad \hspace{3mm} CONSEQ\_PATH p ($\mathcal{SS}^{NO}_{OR}$ p SSp)])  = \\  \hspace{2.5mm} 
	$\prod$ (MAP ($\lambda$ a. 1 - $\prod$ (PROB\_LIST p a)) SSm) \\ 
	$\times$ $\prod$ \\  \hspace{3mm} (MAP \\  \hspace{6mm}  ($\lambda$ b. \\  \hspace{9mm}  (1 -  $\prod$    (PROB\_LIST p  (compl\_list p b)))) SSp)}}
\end{flushleft}

\noindent (C5) Some outcomes of $s$ out of $m$~decisions boxes are NO,  some outcomes of $u$ out of $m$ decisions boxes are~YES and all outcomes of $p$ decisions boxes are NO.
\begin{equation}
	\label{Equation 17} 
	\centering \mathcal{F}_{C5} (t) =   \Bigg(\prod\limits^s_{i=1} 
	\prod\limits^k_{j=2} {\mathcal{F}}_{ij} (t)\Bigg)    \times \Bigg(\prod\limits^u_{i=1}  (1 - 
	\prod\limits^k_{j=2}{\mathcal{F}}_{ij} (t))\Bigg) \times \Bigg(\prod\limits^p_{i=1} 
    (1 - 	\prod\limits^z_{j=2} (1 - {\mathcal{F}}_{ij} (t)))\Bigg)
\end{equation}

\begin{flushleft}
	\texttt{\bf{Theorem 17:}}\\
	\vspace{1pt} 
	\small{\texttt{$\vdash$   prob p \\ \qquad \hspace{1mm}  (CONSEQ\_PATH p \\ \qquad \qquad \hspace{1mm} [CONSEQ\_PATH p ($\mathcal{SS}^{NO}_{AND}$ p SSs); \\ \hspace{3mm} \\ \qquad \qquad \hspace{3mm} CONSEQ\_PATH p ($\mathcal{SS}^{YES}_{AND}$ p SSu); \\ \hspace{3mm} \\ \qquad \qquad \hspace{3mm} CONSEQ\_PATH p ($\mathcal{SS}^{NO}_{OR}$ p SSp)])  = \\  \hspace{2.5mm} 
	$\prod$ (MAP ($\lambda$ a. $\prod$ (PROB\_LIST p a)) SSs) \\ 
	$\times$ $\prod$ (MAP ($\lambda$ b. 1 - $\prod$ (PROB\_LIST p b)) SSu) \\ 
	$\times$ $\prod$ \\  \hspace{3mm} (MAP \\  \hspace{6mm}  ($\lambda$ c. \\  \hspace{9mm}  (1 -  $\prod$    (PROB\_LIST p  (compl\_list p c)))) SSp)}}
\end{flushleft}

\noindent (C6) Some outcomes of $s$ out of $m$~decisions boxes are NO,  some outcomes of $u$ out of $m$ decisions boxes are~YES and all outcomes of $p$ decisions boxes are YES.
\begin{equation}
	\label{Equation 18} 
	 	\centering \mathcal{F}_{C6} (t) =   \Bigg(\prod\limits^s_{i=1} 
	\prod\limits^k_{j=2} {\mathcal{F}}_{ij} (t)\Bigg)    \times \Bigg(\prod\limits^u_{i=1}  (1 - 
	\prod\limits^k_{j=2}{\mathcal{F}}_{ij} (t))\Bigg) \times \Bigg( \prod\limits^p_{i=1} 
	\prod\limits^z_{j=2} (1 - {\mathcal{F}}_{ij} (t))\Bigg)    
	\centering 
\end{equation}

\begin{flushleft}
	\texttt{\bf{Theorem 18:}}\\
	\vspace{1pt} 
	\small{\texttt{$\vdash$   prob p \\ \qquad \hspace{1mm}  (CONSEQ\_PATH p \\ \qquad \qquad \hspace{1mm} [CONSEQ\_PATH p ($\mathcal{SS}^{NO}_{AND}$ p SSs); \\ \hspace{3mm} \\ \qquad \qquad \hspace{3mm} CONSEQ\_PATH p ($\mathcal{SS}^{YES}_{AND}$ p SSu); \\ \hspace{3mm} \\ \qquad \qquad \hspace{3mm} CONSEQ\_PATH p ($\mathcal{SS}^{YES}_{OR}$ p SSp)])  = \\  \hspace{2.5mm} 
	$\prod$ (MAP ($\lambda$ a. $\prod$ (PROB\_LIST p a)) SSs) \\ 
	$\times$ $\prod$ (MAP ($\lambda$ b. 1 - $\prod$ (PROB\_LIST p b)) SSu) \\ 
	$\times$ $\prod$ \\  \hspace{3mm} (MAP \\  \hspace{6mm} ($\lambda$ c. \\  \hspace{9mm}  $\prod$ (PROB\_LIST p (compl\_list p c))) SSp)}}
\end{flushleft}

\noindent (C7) Some outcomes of $s$ out of $p$~decisions boxes are NO,  some outcomes of $u$ out of $p$ decisions boxes are~YES and all outcomes of $m$ decisions boxes are NO.
\begin{equation}
	\label{Equation 19} 
	\centering 
	 	\mathcal{F}_{C7} (t) =   \Bigg(\prod\limits^m_{i=1} 
	\prod\limits^k_{j=2} {\mathcal{F}}_{ij} (t)\Bigg)  \times \Bigg( \prod\limits^u_{i=1} 
	\prod\limits^z_{j=2} (1 - {\mathcal{F}}_{ij} (t))\Bigg) \times   \Bigg(\prod\limits^s_{i=1} 
    (1 - 	\prod\limits^z_{j=2} (1 - {\mathcal{F}}_{ij} (t)))\Bigg)      
\end{equation}

\begin{flushleft}
	\texttt{\bf{Theorem 19:}}\\
	\vspace{1pt} 
	\small{\texttt{$\vdash$   prob p \\ \qquad \hspace{1mm}  (CONSEQ\_PATH p \\ \qquad \qquad \hspace{1mm} [CONSEQ\_PATH p ($\mathcal{SS}^{NO}_{AND}$ p SSm); \\ \hspace{3mm} \\ \qquad \qquad \hspace{3mm} CONSEQ\_PATH p ($\mathcal{SS}^{YES}_{OR}$ p SSu); \\ \hspace{3mm} \\ \qquad \qquad \hspace{3mm} CONSEQ\_PATH p ($\mathcal{SS}^{NO}_{OR}$ p SSs)])  = \\  \hspace{2.5mm} 
	$\prod$ (MAP ($\lambda$ a. $\prod$ (PROB\_LIST p a)) SSm) \\
	$\times$ $\prod$ \\  \hspace{3mm} (MAP \\  \hspace{6mm} ($\lambda$ b. \\  \hspace{9mm}  $\prod$ (PROB\_LIST p (compl\_list p b))) SSu) \\ 
	$\times$ $\prod$ \\  \hspace{3mm} (MAP \\  \hspace{6mm}  ($\lambda$ c. \\  \hspace{9mm}  (1 -  $\prod$    (PROB\_LIST p  (compl\_list p c)))) SSs)}}
\end{flushleft}

\noindent (C8) Some outcomes of $s$ out of $p$~decisions boxes are NO,  some outcomes of $u$ out of $p$ decisions boxes are~YES and all outcomes of $m$ decisions boxes are YES.
\begin{equation}
	\label{Equation 20} 
	\centering 
	 \mathcal{F}_{C8} (t) = \hspace{4mm} \Bigg(\prod\limits^m_{i=1}  (1 - 
	\prod\limits^k_{j=2}{\mathcal{F}}_{ij} (t))\Bigg) \times \Bigg( \prod\limits^u_{i=1} 
	\prod\limits^z_{j=2} (1 - {\mathcal{F}}_{ij} (t))\Bigg) \times 
	\Bigg(\prod\limits^s_{i=1} 
    (1 - 	\prod\limits^z_{j=2} (1 - {\mathcal{F}}_{ij} (t)))\Bigg) 
\end{equation}

\begin{flushleft}
	\texttt{\bf{Theorem 20:}}\\
	\vspace{1pt} 
	\small{\texttt{$\vdash$   prob p \\ \qquad \hspace{1mm}  (CONSEQ\_PATH p \\ \qquad \qquad \hspace{1mm} [CONSEQ\_PATH p ($\mathcal{SS}^{YES}_{AND}$ p SSm); \\ \hspace{3mm} \\ \qquad \qquad \hspace{3mm} CONSEQ\_PATH p ($\mathcal{SS}^{YES}_{OR}$ p SSu); \\ \hspace{3mm} \\ \qquad \qquad \hspace{3mm} CONSEQ\_PATH p ($\mathcal{SS}^{NO}_{OR}$ p SSs)])  = \\  \hspace{2.5mm} 
	$\prod$ (MAP ($\lambda$ a. 1 - $\prod$ (PROB\_LIST p a)) SSm) \\
	$\times$ $\prod$ \\  \hspace{3mm} (MAP \\  \hspace{6mm} ($\lambda$ b. \\  \hspace{9mm}  $\prod$ (PROB\_LIST p (compl\_list p b))) SSu) \\ 
	$\times$ $\prod$ \\  \hspace{3mm} (MAP \\  \hspace{6mm}  ($\lambda$ c. \\  \hspace{9mm}  (1 -  $\prod$    (PROB\_LIST p  (compl\_list p c)))) SSs)}}
\end{flushleft}

\noindent (C9) Some outcomes of $s$ out of $m$~decisions boxes are NO,  some outcomes of $u$ out of $m$ decisions boxes are~YES, some outcomes of $v$ out of $p$~decisions boxes are NO and  some outcomes of $w$ out of $p$ decisions boxes are~YES. 

\begin{equation}
	\label{Equation 21} 
	\centering 
	\begin{split}
	\mathcal{F}_{C9} (t) &=  \Bigg(\prod\limits^s_{i=1} 
	\prod\limits^k_{j=2} {\mathcal{F}}_{ij} (t)\Bigg)   \times \Bigg(\prod\limits^v_{i=1} 
    (1 - 	\prod\limits^z_{j=1} (1 - {\mathcal{F}}_{ij} (t)))\Bigg) \\ &\times \Bigg(\prod\limits^u_{i=1}  (1 - 
	\prod\limits^k_{j=2}{\mathcal{F}}_{ij} (t))\Bigg) \times \Bigg( \prod\limits^w_{i=1} 
	\prod\limits^z_{j=2} (1 - {\mathcal{F}}_{ij} (t))\Bigg)  
	\end{split}	
\end{equation}

\begin{flushleft}
	\texttt{\bf{Theorem 21:}}\\
	\vspace{1pt} 
	\small{\texttt{$\vdash$   prob p \\ \qquad \hspace{1mm}  (CONSEQ\_PATH p \\ \qquad \qquad \hspace{1mm} [CONSEQ\_PATH p ($\mathcal{SS}^{NO}_{AND}$ p SSs); \\ \hspace{3mm} \\ \qquad \qquad \hspace{3mm} CONSEQ\_PATH p ($\mathcal{SS}^{YES}_{AND}$ p SSu); \\ \hspace{3mm} \\ \qquad \qquad \hspace{3mm} CONSEQ\_PATH p ($\mathcal{SS}^{NO}_{OR}$ p SSv); \\ \hspace{3mm} \\ \qquad \qquad \hspace{3mm} CONSEQ\_PATH p ($\mathcal{SS}^{YES}_{OR}$ p SSw)])  = \\  \hspace{2.5mm} 
	$\prod$ (MAP ($\lambda$ a. $\prod$ (PROB\_LIST p a)) SSs)  \\
	$\times$ $\prod$ (MAP ($\lambda$ b. 1 - $\prod$ (PROB\_LIST p b)) SSu) \\ 
	$\times$ $\prod$ \\  \hspace{3mm} (MAP \\  \hspace{6mm}  ($\lambda$ c. \\  \hspace{9mm}  (1 -  $\prod$    (PROB\_LIST p  (compl\_list p c)))) SSv) \\
	$\times$ $\prod$ \\  \hspace{3mm} (MAP \\  \hspace{6mm} ($\lambda$ d. \\  \hspace{9mm}  $\prod$ (PROB\_LIST p (compl\_list p d))) SSw)   }}
\end{flushleft}

Therefore, by verifying all the above-mentioned theorems in HOL4, we showed the completeness  of  our proposed formal approach and thereupon solving the scalability problem of CCD analysis for any given large engineering complex system at the subsystem level \cite{Etree_tp}. \\

\textit{Property 7}: A \textit{generic}  probabilistic expression of \texttt{CONSEQ\_BOX} for a certain event occurrence in the entire system as the sum of all individual probabilities of all $\mathcal{M}$~\texttt{CONSEQ\_PATH} ending with that event:

\begin{flushleft}
\label{Theorem 8}
	\texttt{\bf{Theorem 22:}}\\
	\vspace{1pt} 
	\small{\texttt{$\vdash$ \small{\texttt{Let}} \\ \quad CONSEQ\_PATHS $\mathrm{L}_\mathcal{M}$ = MAP ($\lambda$a. CONSEQ\_PATH p a) $\mathrm{L}_\mathcal{M}$) \\ \quad in \\ \quad prob\_space p $\wedge$ MUTUAL\_INDEP p $\mathrm{L}_\mathcal{M}$  $\wedge$ \\ \quad disjoint (CONSEQ\_PATHS  $\mathrm{L}_\mathcal{M}$) $\wedge$   ALL\_DISTINCT (CONSEQ\_PATHS $\mathrm{L}_\mathcal{M}$)  $\Rightarrow$ \\  \hspace{2.5mm}  prob p (CONSEQ\_BOX p $\mathrm{L}_\mathcal{M}$) = $\sum$  (PROB\_LIST p (CONSEQ\_PATHS $\mathrm{L}_\mathcal{M}$))}}
\end{flushleft}

\noindent where the HOL4 function \texttt{disjoint} ensures that each pair of elements in a given list is mutually exclusive while the function \texttt{ALL\_DISTINCT}  ensures that each pair is distinct. The function $\sum$ is defined to sum the events for a given list. Remark that all above-mentioned CCD new formulations have been  \textit{formally verified} in HOL4, where the proof-script amounts to about 16,000 lines of HOL4 code, which can be downloaded for use from~\cite{Etree_tp}. Also, this code can be extended, with some basic knowhow about HOL4, to perform dynamic failure analysis of dynamic subsystems where no dependencies exist in subsystems using DFTs, such as PAND and~SP,~i.e,~CCD~reliability analysis of~\textit{Type~II}~(see Fig.~\ref{Fig: CCD analysis Type B}).

To illustrate the applicability of our proposed approach, in the next section, we present the formal CCD step-analysis of  the standard IEEE 39-bus electrical power network and verify its reliability indexes ($\mathcal{FOR}$ and $\mathcal{SAIDI}$), which are commonly used as reliability indicators by electric power~utilities. 

\section{Electrical Power 39-bus Network System}
\label{Electrical Power 39-bus Network System}
An electrical power network is an interconnected grid for delivering electricity from producers to customers. The power network system consists of three main zones \cite{fang2011smart}:~(i)~generating stations that produce electric power; (ii)  transmission lines that carry power from sources to loads; and (iii)~distribution lines that connect individual consumers. Due to the complex and integrated nature of the power network, failures in any zone of the system can cause widespread catastrophic disruption of supply~\cite{fang2011smart}. Therefore a rigorous formal cause-consequence analysis of the grid is essential in order to reduce the risk situation of a blackout and enable back-up decisions~\cite{allan2013reliability}.  For power network safety assessment, reliability engineers have been dividing the power network into three main hierarchical levels~\cite{vcepin2011assessment}: (a)~generation systems; (b)~composite generation and transmission (or bulk power) systems; and (c)~distribution systems. We can use our proposed CCD formalization for the formal modeling and analysis of any hierarchical level in the power network. In this case study, we focus on the generation part only, i.e., hierarchical level I. Also, we can evaluate the Force Outage Rate~($\mathcal{FOR}$) for the generation stations, which is defined as the probability of the unit unavailability to produce power due to unexpected equipment failure~\cite{allan2013reliability}. Additionally, we can determine the System Average Interruption Duration Index~($\mathcal{SAIDI}$), which is used to indicate the average duration for each customer served to experience a sustained outage. $\mathcal{SAIDI}$ is defined as the sum of all customer interruption durations (probability of load failures $\mbox{\Lightning}$ multiplying by  the mean-time-to-repair  the failures and the number of customers that are affected by these failures) over the total number of customers~served~\cite{allan2013reliability}:
\begin{equation}
	\label{Eq:SAIDI}
	\centering 
	\mathcal{SAIDI}  = \frac{\sum_{{\mathcal{P}} (\mathcal{X}_{\mbox{\Lightning}}) \times \mathrm{MTTR}_{\mathcal{X}} \times \mathrm{CN}_\mathcal{X}}}{\sum_{\mathrm{CN}_\mathcal{X}}}
\end{equation}

\noindent where $\mathrm{CN}_\mathcal{X}$ is the number of customers for a certain location \textit{$\mathcal{X}$} while $\mathrm{MTTR}_{\mathcal{X}}$ is the mean-time-to-repair the failure that occurred at \textit{$\mathcal{X}$}. We formally  define a function $\sum\nolimits^T_{\mbox{\Lightning}}$ in HOL4 to sum all customer interruption durations. Also, we formally define a generic function $\mathcal{SAIDI}$  by dividing the output of $\sum\nolimits^T_{\mbox{\Lightning}}$ over the total number of customers served, in HOL4 as:

\begin{flushleft}
\label{Definition 7}
	\texttt{\bf{Definition 10:}}\\
	\vspace{1pt} \small{\texttt{$\vdash$  $\sum\nolimits^T_{\mbox{\Lightning}}$ ($\mathrm{\small{\texttt{L}}}$::$\mathrm{\small{\texttt{L}}}_\mathcal{M}$)  (MTTR::$\mathrm{\small{\texttt{MTTR}}}_\mathcal{M}$) (CN:$\mathrm{\small{\texttt{CN}}}_\mathcal{M}$) p = \\ \quad 
   prob p (CONSEQ\_BOX p $\mathrm{\small{\texttt{L}}}_\mathcal{M}$) $\times$ MTTR  $\times$ CN  +  $\sum\nolimits^T_{\mbox{\Lightning}}$ $\mathrm{\small{\texttt{L}}}_\mathcal{M}$ $\mathrm{\small{\texttt{MTTR}}}_\mathcal{M}$ $\mathrm{\small{\texttt{CN}}}_\mathcal{M}$ p}}
\end{flushleft}
\begin{flushleft}
	\texttt{\bf{Definition 11:}}\\
  \vspace{1pt} \small{\texttt{$\vdash$ 
     $\mathcal{SAIDI}$  $\mathrm{\small{\texttt{L}}}_\mathcal{M}$ $\mathrm{\small{\texttt{MTTR}}}_\mathcal{M}$ $\mathrm{\small{\texttt{CN}}}_\mathcal{M}$ p = 
    $\dfrac {\sum\nolimits^T_{\mbox{\Lightning}} \mathrm{\small{\texttt{L}}}_\mathcal{M} \hspace{1mm} \mathrm{\small{\texttt{MTTR}}}_\mathcal{M} \hspace{1mm} \mathrm{\small{\texttt{CN}}}_\mathcal{M} \hspace{1mm} p} {\sum \mathrm{\small{\texttt{CN}}}_\mathcal{M}}$}}
\end{flushleft}

\noindent where \texttt{$\mathrm{\small{\texttt{L}}}_\mathcal{M}$} is the list of CCD  paths, \texttt{$\mathrm{\small{\texttt{MTTR}}}_\mathcal{M}$} is the list of meantime to repairs, and \texttt{$\mathrm{\small{\texttt{CN}}}_\mathcal{M}$} is the list of customer~numbers. The function $\sum\nolimits^T_{\mbox{\Lightning}}$ (Definition 10) models the numerator of Eq.~\ref{Eq:SAIDI}, which is the sum of all customer interruption durations at different locations in the electrical power grid. Each probability of failure is obtained by evaluating a \texttt{CONSEQ\_BOX} consisting of a list of $\mathcal{M}$ \texttt{CONSEQ\_PATH}, which cause that failure. Definition~11 represents the division of output of Definition~10 over the total number of customers at all those locations as~described~in~Eq.~\ref{Eq:SAIDI}. \\

\begin{figure} [!t]
	\begin{center}
		\centering
		\includegraphics[width= 1 \columnwidth]{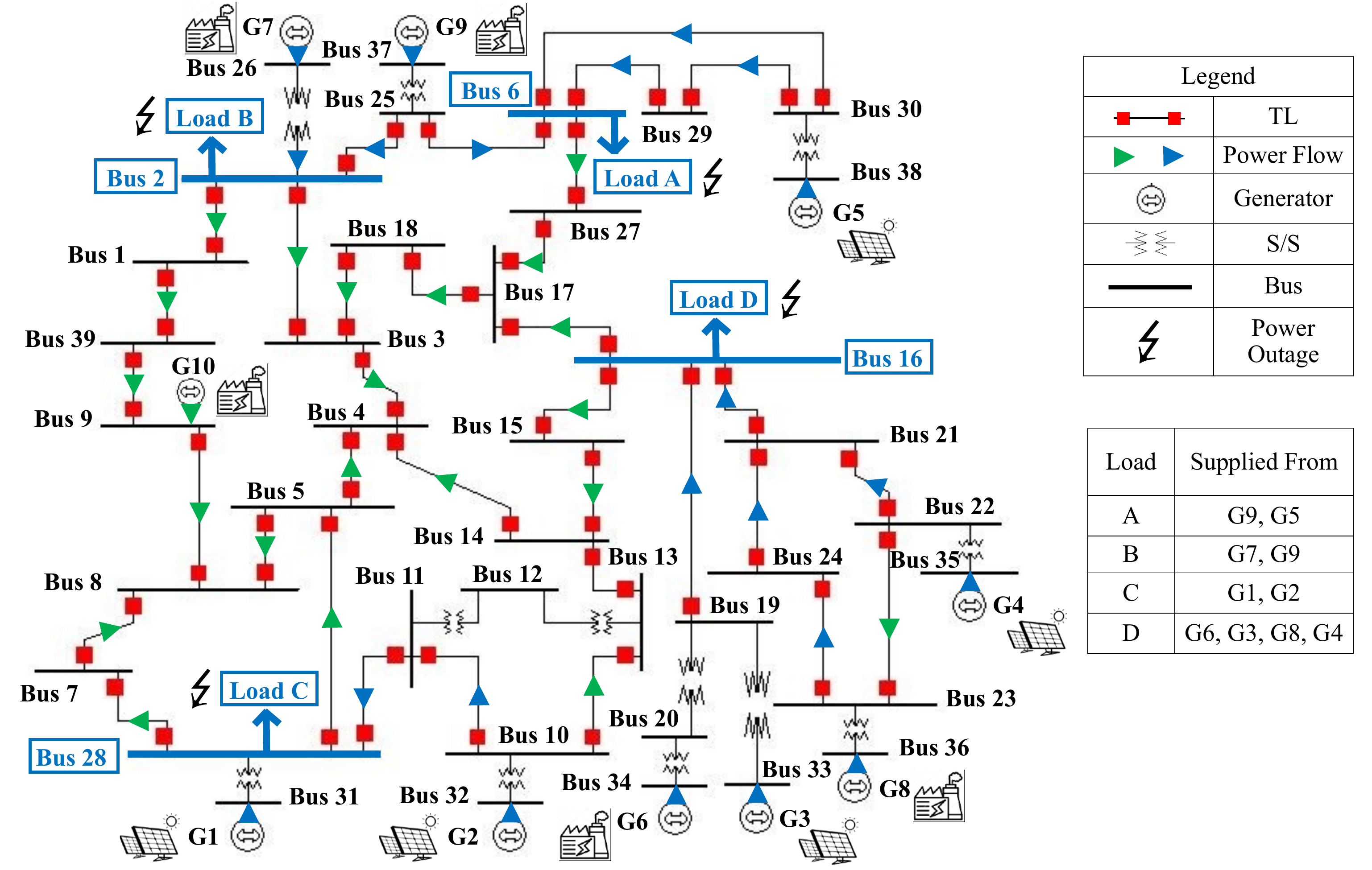} 
	\end{center}
	\caption{IEEE 39-bus Electrical Power Network~\cite{bhatt2017analysis}}
	\label{Fig: Case study}
\end{figure}

Consider a standard \textit{IEEE 39-bus} electrical power network test system consisting of 10 generators (G), 12 substations~(S/S), 39 Buses (Bus), and 34 transmission lines~(TL), as shown in Fig.~\ref{Fig: Case study}~\cite{bhatt2017analysis}. Assuming the generators G1-G10 are of two types: (i)  solar photo-voltaic (PV) power plants G1-G5; and (ii) steam power plants G6-G10. Using the Optimal Power Flow (OPF) optimization \cite{gan2000stability}, we can determine the flow of electricity from generators to consumers in the power network. Typically, we are only interested in evaluating the duration of certain failure events occurrence for specific loads in the grid.  For instance, if we consider the failure of load~A, which according to the OPF is supplied from G9 and G5 only, as shown in Fig.~\ref{Fig: Case study}, then the failure of either one or both power plants will lead to a partial or a complete blackout failure at that load, respectively. Assuming the failure of two consecutive power plants causes  a complete blackout of the load. Hence, considering the disruption cases of \textit{only one} supply generator, then different partial failures for loads A, B, C and D, as shown in Fig.~\ref{Fig: Case study}, can be obtained by observing different failures in the power network~as:
\begin{enumerate}[a.]
	\item $\begin{aligned}[t] 
	\mathcal{P}(\mathrm{Load}_{A\mbox{\Lightning}}) = 
	 & (1 - \mathcal{FOR}_{G_9})  \times \mathcal{FOR}_{G_5}  +   \mathcal{FOR}_{G_9} \times (1 - \mathcal{FOR}_{G_5}) 
	\end{aligned}$
	
	\item $\begin{aligned}[t] 
	\mathcal{P}(\mathrm{Load}_{B\mbox{\Lightning}}) = 
	 & (1 - \mathcal{FOR}_{G_7})  \times \mathcal{FOR}_{G_9}  +   \mathcal{FOR}_{G_7}  \times (1 - \mathcal{FOR}_{G_9})
	\end{aligned}$
	
	\item $\begin{aligned}[t] 
	\mathcal{P}(\mathrm{Load}_{C\mbox{\Lightning}}) = 
	 & (1 - \mathcal{FOR}_{G_1})  \times \mathcal{FOR}_{G_2}  +   \mathcal{FOR}_{G_1} \times (1 - \mathcal{FOR}_{G_2})
	\end{aligned}$
	
	\item $\begin{aligned}[t]
    \mathcal{P}(\mathrm{Load}_{D\mbox{\Lightning}}) &  = 
       (1 - \mathcal{FOR}_{G_6}) \times (1 - \mathcal{FOR}_{G_3}) \times
     (1 - \mathcal{FOR}_{G_8}) \times \mathcal{FOR}_{G_4}
	\\ & \hspace{0.7mm} + (1 - \mathcal{FOR}_{G_6}) \times (1 - \mathcal{FOR}_{G_3}) \times
    \mathcal{FOR}_{G_8} \times (1 - \mathcal{FOR}_{G_4}) 
	\\ & \hspace{0.7mm} + (1 - \mathcal{FOR}_{G_6}) \times \mathcal{FOR}_{G_3} \times
   (1 - \mathcal{FOR}_{G_8}) \times (1 - \mathcal{FOR}_{G_4}) 
	\\ & \hspace{0.7mm} +  \mathcal{FOR}_{G_6}  \times (1 - \mathcal{FOR}_{G_3}) \times
    (1 - \mathcal{FOR}_{G_8}) \times (1 - \mathcal{FOR}_{G_4}) 
	 \end{aligned}$
\end{enumerate}

\noindent Therefore, the assessment of $\mathcal{SAIDI}$ for the Grid (G) shown in Fig.~\ref{Fig: Case study}, including an evaluation for the $\mathcal{FOR}$ of all its power plants, can be written mathematically as:
\begin{equation}
	\label{Eq:probability}
	\centering 
	\mathcal{SAIDI}_{G} = \frac{\mathcal{P}(\mathrm{Load}_{A\mbox{\Lightning}}) \times \mathrm{MTTR}_{\mathrm{Load}_{A}} \times \mathrm{CN}_{\mathrm{Load}_{A}}  + \dots} {\mathrm{CN}_{\mathrm{Load}_{A}} + \mathrm{CN}_{\mathrm{Load}_{B}} + \mathrm{CN}_{\mathrm{Load}_{C}} + \mathrm{CN}_{\mathrm{Load}_{D}}} 
\end{equation}

\subsection{Formal CCD Analysis in HOL4}
We can apply our \textit{four} steps of CCD formalization to verify the expression of  $\mathcal{SAIDI}$ in terms of the power plant generator components, in HOL4 as:\\ 

\noindent\textit{Step 1 (Component failure events)}:\\
The schematic FT models of a typically PV power plant consisting of 2 solar farms~\cite{alferidi2017development} and a steam power plant consisting of 3 generators \cite{allan2013reliability} are shown in Fig.~\ref{FT Model of a typical PV power plant} and  Fig.~\ref{Fig: FT Model of a typical steam power plant}, respectively. Using the formal FT  modeling, we can formally define the FT models of both plants, in HOL4~as:

\begin{figure}
\centering
\parbox{5cm}{
\includegraphics[width=6cm]{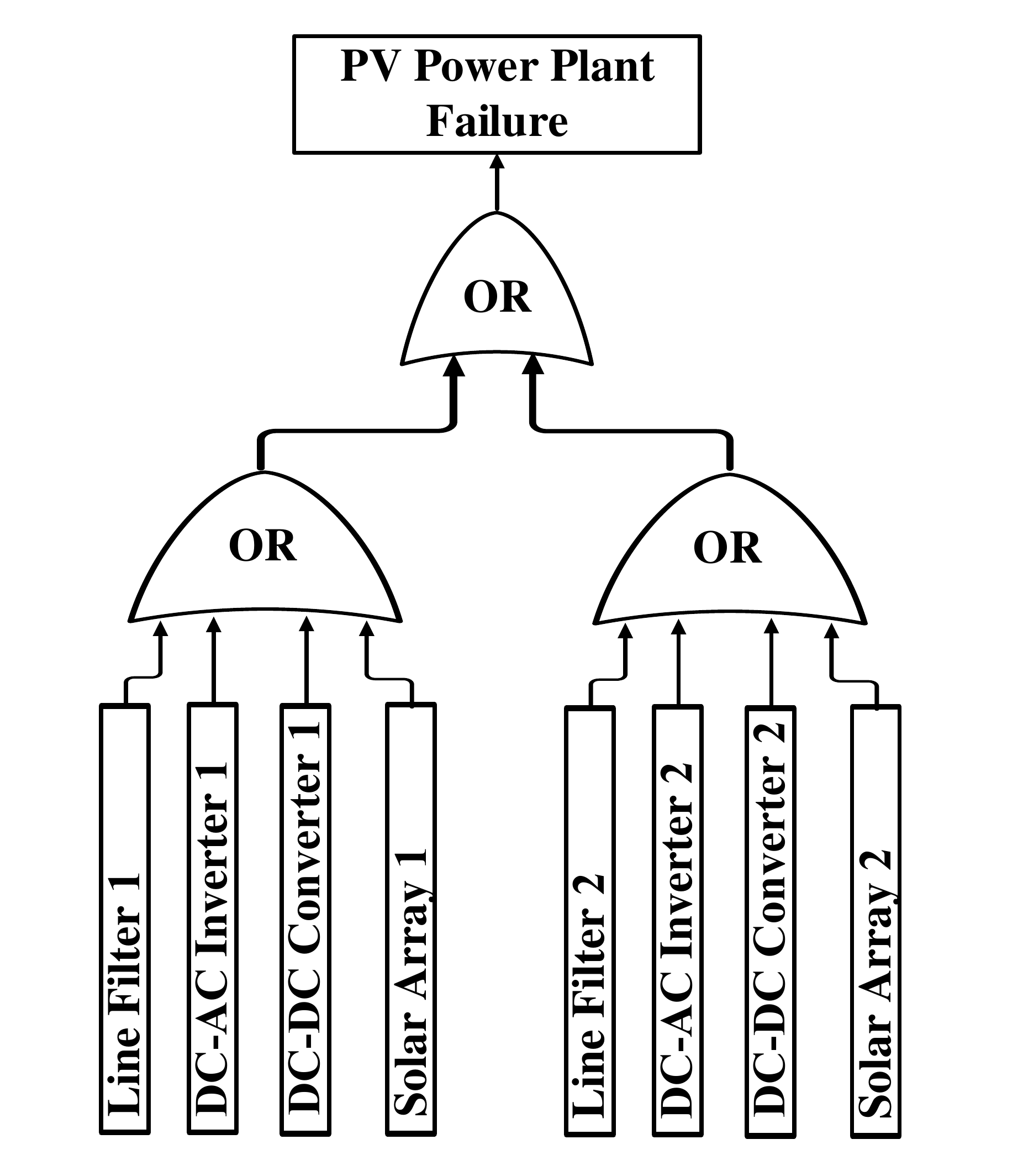}
\caption{FT Model of a PV Power Plant}
\label{FT Model of a typical PV power plant}}
\qquad
\begin{minipage}{5cm}
\includegraphics[width=6cm]{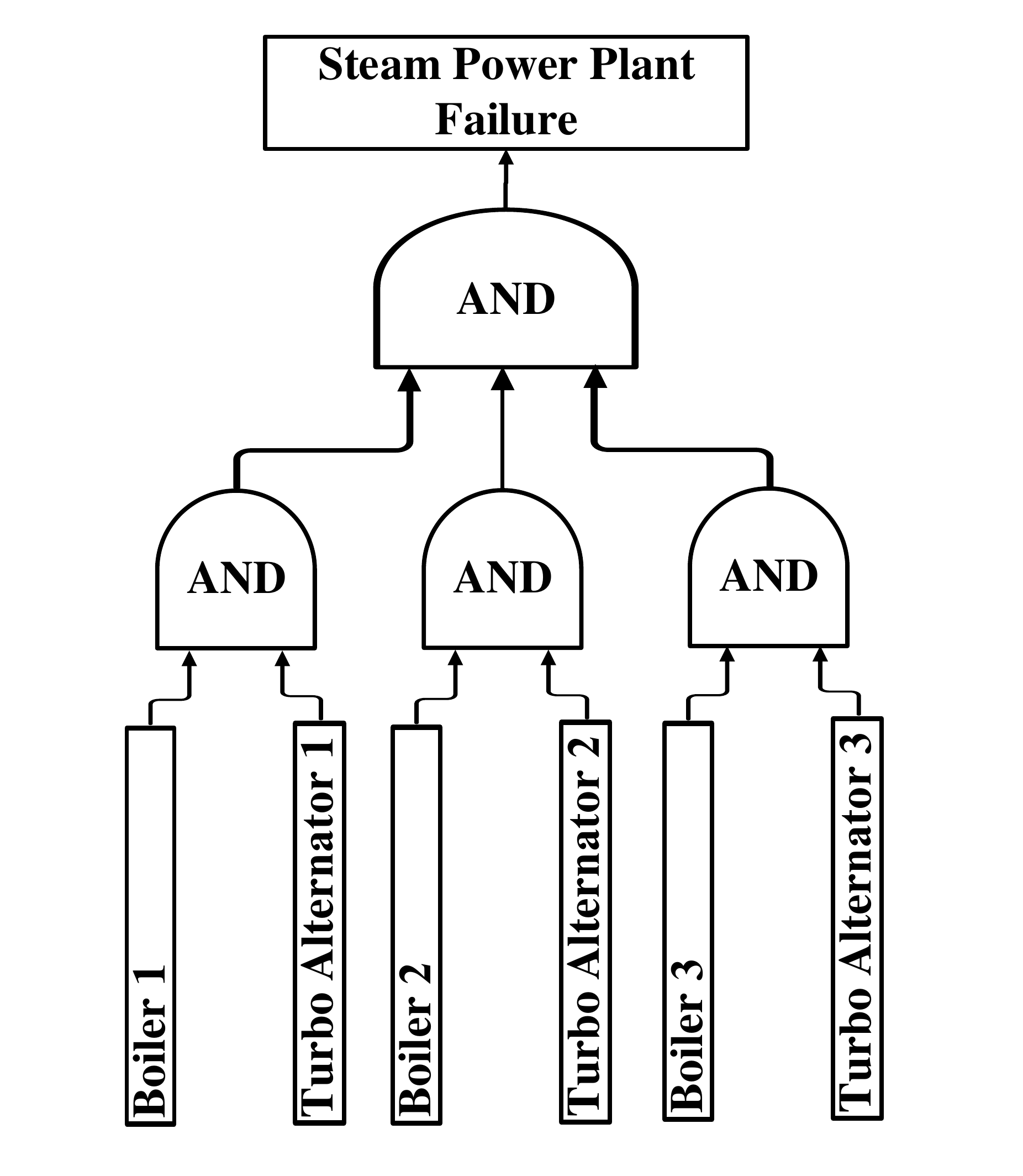}
\caption{FT Model of a Steam Power Plant}
\label{Fig: FT Model of a typical steam power plant}
\end{minipage}
\end{figure}

\begin{flushleft}
	\texttt{\bf{Definition 12:}}\\
  \vspace{1pt} \small{\texttt{$\vdash$ 
     $\mathrm{FT}_{PV}$  p [LF1;LF2] [DC\_DC1;DC\_DC2]  [SA1;SA2]  [DC\_AC1;DC\_AC2] =  \\ \hspace{2mm}
     FTree p (OR [OR [LF1;DC\_DC1;DC\_AC1;SA1];  OR [LF2;DC\_DC2;DC\_AC2;SA2]])}}
\end{flushleft}

\vspace{-2mm}

\begin{flushleft}
	\texttt{\bf{Definition 13:}}\\
  \vspace{1pt} \small{\texttt{$\vdash$ 
$\mathrm{FT}_{STEAM}$  p  [BO1;BO2;BO3] [TA1;TA2;TA3] = \\ \hspace{2mm}
FTree p (AND [AND [BO1;TA1];  AND [BO2;TA2];  AND [BO3;TA3]])}}
\end{flushleft}

\begin{figure}[!t]
	\includegraphics[width= 1 \columnwidth]{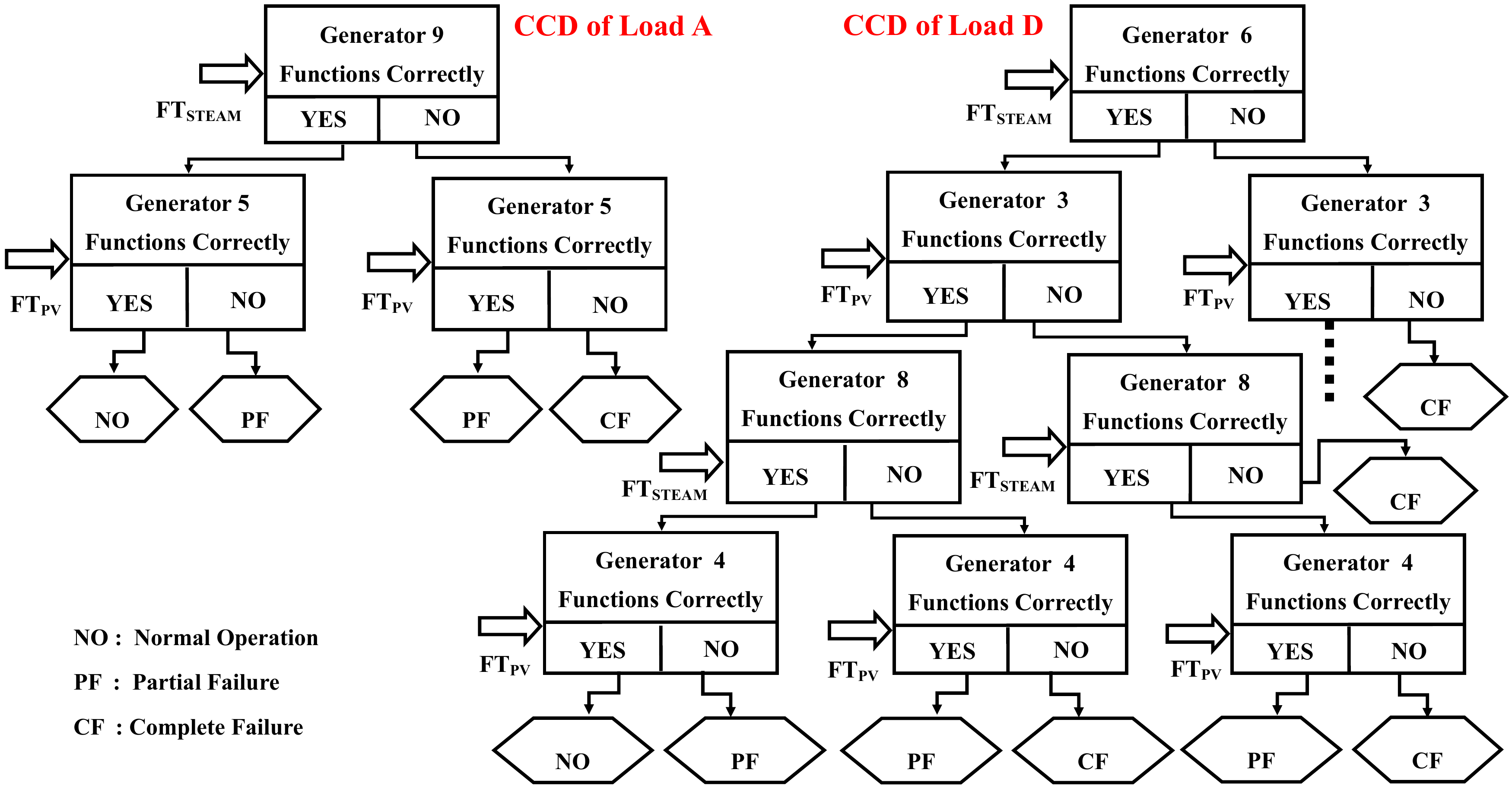} 
	\centering
	\caption{CCD Analysis of Loads A and D }
	\label{Fig: CCD Analysis of Loads A and D in the Electrical Power Network System}
\end{figure}

\noindent\textit{Steps 2 and 3 (Construction of a CCD and Reduction)}:\\
Construct a formal complete CCD for all loads in our case study (Fig.~\ref{Fig: Case study}), i.e.,  A, B, C, and D, then remove the irrelevant decision boxes according to the electrical power network functional behavior. For instance, we can model the CCD models for loads A and D, as shown in Fig. \ref{Fig: CCD Analysis of Loads A and D in the Electrical Power Network System}, respectively, in HOL4~as:

\begin{flushleft}
	\texttt{\bf{Definition 14:}}\\
	\vspace{1pt} 
	\small{\texttt{$\vdash$  {CCD\_LOAD\_A} =   \\ \quad
	CONSEQ\_BOX p \\ \quad  \hspace{2mm} [[DEC\_BOX p 1 ($\overline{\mathrm{FT}_{STEAM}}$,$\mathrm{FT}_{STEAM}$);DEC\_BOX p 1 ($\overline{\mathrm{FT}_{PV}}$,$\mathrm{FT}_{PV}$)]; \\ \quad \quad \hspace{0.5mm}
	[DEC\_BOX p 1 ($\overline{\mathrm{FT}_{STEAM}}$,$\mathrm{FT}_{STEAM}$);DEC\_BOX p 0 ($\overline{\mathrm{FT}_{PV}}$,$\mathrm{FT}_{PV}$)];\\ \quad \quad \hspace{0.5mm} 
	[DEC\_BOX p 0 ($\overline{\mathrm{FT}_{STEAM}}$,$\mathrm{FT}_{STEAM}$);DEC\_BOX p 1 ($\overline{\mathrm{FT}_{PV}}$,$\mathrm{FT}_{PV}$)];\\ \quad \quad \hspace{0.5mm}
	[DEC\_BOX p 0 ($\overline{\mathrm{FT}_{STEAM}}$,$\mathrm{FT}_{STEAM}$);DEC\_BOX p 0 ($\overline{\mathrm{FT}_{PV}}$,$\mathrm{FT}_{PV}$)]]}}
\end{flushleft}

\begin{flushleft}
	\texttt{\bf{Definition 15:}}\\
	\vspace{1pt} 
	\small{\texttt{$\vdash$  {CCD\_LOAD\_D} =    \\ \quad
	CONSEQ\_BOX p \\ \quad \hspace{1.5mm}  [[DEC\_BOX p 1 ($\overline{\mathrm{FT}_{STEAM}}$,$\mathrm{FT}_{STEAM}$);DEC\_BOX p 1 ($\overline{\mathrm{FT}_{PV}}$,$\mathrm{FT}_{PV}$); \\ \quad \quad \hspace{1.6mm} DEC\_BOX p 1 ($\overline{\mathrm{FT}_{STEAM}}$,$\mathrm{FT}_{STEAM}$);DEC\_BOX p 1 ($\overline{\mathrm{FT}_{PV}}$,$\mathrm{FT}_{PV}$)]; \\ \quad \hspace{4mm}  [DEC\_BOX p 1 ($\overline{\mathrm{FT}_{STEAM}}$,$\mathrm{FT}_{STEAM}$);DEC\_BOX p 1 ($\overline{\mathrm{FT}_{PV}}$,$\mathrm{FT}_{PV}$); \\ \quad \quad \hspace{2.2mm} DEC\_BOX p 1 ($\overline{\mathrm{FT}_{STEAM}}$,$\mathrm{FT}_{STEAM}$);DEC\_BOX p 0 ($\overline{\mathrm{FT}_{PV}}$,$\mathrm{FT}_{PV}$)];   \\ \qquad  \qquad  \qquad \qquad \hspace{0.5mm} \vdots  \\ \quad \hspace{4mm} [DEC\_BOX p 0 ($\overline{\mathrm{FT}_{STEAM}}$,$\mathrm{FT}_{STEAM}$);DEC\_BOX p 0 ($\overline{\mathrm{FT}_{PV}}$,$\mathrm{FT}_{PV}$)]]}}
\end{flushleft}

\noindent\textit{Step 4 (Probabilistic analysis)}:\\
We can use our proposed formal approach to express subsystem-level failure/reliability probabilistic expressions of electrical power grids, which enable us to analyze the cascading dependencies with many subsystem levels, based on any probabilistic distribution. In this work, we assumed that the failure of each component is exponentially distributed (i.e., \texttt{CDF p X t = 1 $-$ $\mathrm{e}^{(-\lambda_{X} t)}$},  where $\lambda_{X}$ is the failure rate of the variable \textit{X} and \textit{t} is a time~index).  

\subsubsection{$\mathcal{FOR}$ Analysis}
Using Definitions 12 and 13  with the assumption that the failure states of components are exponentially distributed, we can formally specify the probabilistic $\mathcal{FOR}$ expression for both PV and steam power plants, in HOL4~as:

\begin{flushleft}
	\texttt{\bf{Definition 16:}}\\
  \vspace{1pt} \small{\texttt{$\vdash$ $\mathcal{FOR}_{PV}$ p [LF1;LF2] [DC\_DC1;DC\_DC2]  [SA1;SA2]  [DC\_AC1;DC\_AC2] = \\ \hspace{2mm} prob p ($\mathrm{FT}_{PV}$  p ($\downarrow$  [LF1;LF2]) ($\downarrow$ [DC\_DC1;DC\_DC2])  \\ \qquad  \qquad \qquad \qquad \hspace{2mm} ($\downarrow$  [SA1;SA2]) ($\downarrow$  [DC\_AC1;DC\_AC2]))}}
\end{flushleft}

\vspace{-2mm}

\begin{flushleft}
	\texttt{\bf{Definition 17:}}\\
  \vspace{1pt} \small{\texttt{$\vdash$ $\mathcal{FOR}_{STEAM}$ p  [BO1;BO2;BO3] [TA1;TA2;TA3] = \\ \hspace{2mm}
prob p ($\mathrm{FT}_{STEAM}$  p  ($\downarrow$ [BO1;BO2;BO3]) ($\downarrow$ [TA1;TA2;TA3]) }}
\end{flushleft}

\noindent where the function $\downarrow$ takes a list of $\mathcal{N}$ components and assigns an exponential failing event to each component in the list. \\
\vspace{-2mm}

We can formally \textit{verify} the above-expressions of $\mathcal{FOR}_{PV}$ and $\mathcal{FOR}_{STEAM}$, in HOL4~as: 

\begin{flushleft}
	\texttt{\bf{Theorem 23:}}\\
  \vspace{1pt} \small{\texttt{$\vdash$ $\mathcal{FOR}_{PV}$ p [LF1;LF2] [DC\_DC1;DC\_DC2] [SA1;SA2]  [DC\_AC1;DC\_AC2] = \\ \hspace{2mm}  $ 1 - \mathrm{e}^{(-\lambda_{LF1} t)}$  $\times$ $\mathrm{e}^{(-\lambda_{LF2} t)}$  $\times$ 
           $\mathrm{e}^{(-\lambda_{DC\_DC1} t)}$  $\times$  
           $\mathrm{e}^{(-\lambda_{DC\_DC2} t)}$  $\times$ $\mathrm{e}^{(-\lambda_{SA1} t)}$  $\times$ 
           $\mathrm{e}^{(-\lambda_{SA2} t)}$  $\times$ 
          \\ \hspace{9mm}  $\mathrm{e}^{(-\lambda_{DC\_AC1} t)}$  $\times$ 
           $\mathrm{e}^{(-\lambda_{DC\_AC2} t)}$
           }}
\end{flushleft}

\vspace{-4mm}

\begin{flushleft}
	\texttt{\bf{Theorem 24:}}\\
  \vspace{1pt} \small{\texttt{$\vdash$ $\mathcal{FOR}_{STEAM}$ p  [BO1;BO2;BO3]  [TA1;TA2;TA3] =  \\
               \hspace{2mm} $(1 - \mathrm{e}^{(-\lambda_{BO1} t)})$  $\times$ 
           $(1 - \mathrm{e}^{(-\lambda_{BO2} t)})$  $\times$ 
           $(1 - \mathrm{e}^{(-\lambda_{BO3} t)})$  $\times$  
           $(1 - \mathrm{e}^{(-\lambda_{TA1} t)})$  $\times$  \\
               \hspace{2mm}
           $(1 - \mathrm{e}^{(-\lambda_{TA2} t)})$  $\times$  
           $(1 - \mathrm{e}^{(-\lambda_{TA3} t)})$
           }}
\end{flushleft}

\subsubsection{$\mathcal{SAIDI}$ Analysis}

Using Theorems 1-24 with the assumption that the failure states of components are exponentially distributed,  we can formally verify $\mathcal{SAIDI}_G$ (Eq. \ref{Eq:probability}), in HOL4~as:

\vspace{-2mm}

\begin{flushleft}
	\texttt{\bf{Theorem 25:}}\\
	\vspace{1pt} 
	\small{\texttt{$\vdash$ $\mathcal{SAIDI}$ \\ \hspace{-1mm}
[[CONSEQ\_PATH p \\ \quad 
            [DEC\_BOX p 1 \\ \qquad \hspace{0.2mm}
               (FTree p (NOT ($\mathrm{FT}_{STEAM}$ p ($\downarrow$ [BO1;BO2;BO3])    ($\downarrow$ [TA1;TA2;TA3]))), \\ \qquad  \qquad \qquad \qquad \qquad 
	       $\mathrm{FT}_{STEAM}$ p ($\downarrow$ [BO1;BO2;BO3])
	       ($\downarrow$ [TA1;TA2;TA3])); \\ \hspace{4mm}
             DEC\_BOX p 0 \\ \qquad \hspace{0.2mm}
               (FTree p (NOT ($\mathrm{FT}_{PV}$ p ($\downarrow$  [LF1;LF2])  ($\downarrow$ [DC\_DC1;DC\_DC2])   \\ \qquad  \qquad \quad \qquad \qquad \qquad \qquad \hspace{2.5mm}  ($\downarrow$  [SA1;SA2]) 
							 ($\downarrow$  [DC\_AC1;DC\_AC2]))),\\ \qquad  \qquad \qquad \qquad \qquad 
	          $\mathrm{FT}_{PV}$ p ($\downarrow$ [LF1;LF2])   ($\downarrow$ [DC\_DC1;DC\_DC2]) \\ \qquad  \qquad \quad \qquad \qquad \qquad \qquad \hspace{2.5mm}  ($\downarrow$ [SA1;SA2]) ($\downarrow$ [DC\_AC1;DC\_AC2]))];  \\ \quad
            [DEC\_BOX p 0 \\ \qquad \hspace{0.2mm}
               (FTree p (NOT ($\mathrm{FT}_{STEAM}$ p ($\downarrow$ [BO1;BO2;BO3])     ($\downarrow$ [TA1;TA2;TA3]))), \\ \qquad  \qquad \qquad \qquad \qquad 
	       $\mathrm{FT}_{STEAM}$ p ($\downarrow$ [BO1;BO2;BO3]) 
	       ($\downarrow$ [TA1;TA2;TA3])); \\ \hspace{4mm}
             DEC\_BOX p 1 \\ \qquad \hspace{0.2mm}
               (FTree p (NOT ($\mathrm{FT}_{PV}$ p ($\downarrow$  [LF1;LF2])  ($\downarrow$ [DC\_DC1;DC\_DC2])   \\ \quad  \qquad \qquad \qquad \qquad \qquad \qquad \hspace{2.7mm}  ($\downarrow$  [SA1;SA2]) 
							 ($\downarrow$  [DC\_AC1;DC\_AC2]))),\\ \qquad  \qquad \qquad \qquad \qquad 
	          $\mathrm{FT}_{PV}$ p ($\downarrow$ [LF1;LF2])   ($\downarrow$ [DC\_DC1;DC\_DC2]) \\ \qquad  \qquad \quad \qquad \qquad \qquad \qquad \hspace{2.5mm}  ($\downarrow$ [SA1;SA2]) ($\downarrow$ [DC\_AC1;DC\_AC2]))]]; 
	          \\ \hspace{2mm} \dots] \\ \hspace{-1mm}[MTTR\_LoadA;MTTR\_LoadB;MTTR\_LoadC;MTTR\_LoadD] \\ \hspace{-1mm}[CN\_LoadA; CN\_LoadB; CN\_LoadC; CN\_LoadD] p = 
$$\frac{\begin{aligned}	          
         &((1 - (1 - \mathrm{e}^{(-\lambda_{BO1} t)}) \times
               (1 - \mathrm{e}^{(-\lambda_{BO2} t)})  \times
               (1 - \mathrm{e}^{(-\lambda_{BO3} t)})  \times  \\
               &\hspace{1.9mm} \quad \quad (1 - \mathrm{e}^{(-\lambda_{TA1} t)}) \times 
               (1 - \mathrm{e}^{(-\lambda_{TA2} t)})  \times 
               (1 - \mathrm{e}^{(-\lambda_{TA3} t)}))  \times  \\
               &\hspace{1.5mm}
          (1 - \mathrm{e}^{(-\lambda_{LF1} t)}  \times 
               \mathrm{e}^{(-\lambda_{LF2} t)}  \times
               \mathrm{e}^{(-\lambda_{DC\_DC1} t)}   \times  
               \mathrm{e}^{(-\lambda_{DC\_DC2} t)}  \times \\
               &\hspace{1.9mm}  \quad \quad 
               \mathrm{e}^{(-\lambda_{DC\_AC1} t)} \times 
               \mathrm{e}^{(-\lambda_{DC\_AC2} t)}  \times 
               \mathrm{e}^{(-\lambda_{SA1} t)}   \times  
               \mathrm{e}^{(-\lambda_{SA2} t)}) +  \\
               &\hspace{1.5mm} 
           (1 - \mathrm{e}^{(-\lambda_{BO1} t)})  \times 
           (1 - \mathrm{e}^{(-\lambda_{BO2} t)})  \times 
           (1 - \mathrm{e}^{(-\lambda_{BO3} t)})  \times  \\
               &\hspace{1.5mm}
           (1 - \mathrm{e}^{(-\lambda_{TA1} t)})  \times 
           (1 - \mathrm{e}^{(-\lambda_{TA2} t)})  \times 
           (1 - \mathrm{e}^{(-\lambda_{TA3} t)}) \times  \\
               &\hspace{1.5mm}
           \mathrm{e}^{(-\lambda_{LF1} t)}  \times 
           \mathrm{e}^{(-\lambda_{LF2} t)}  \times 
           \mathrm{e}^{(-\lambda_{DC\_DC1} t)}  \times 
           \mathrm{e}^{(-\lambda_{DC\_DC2} t)}  \times  \\
               &\hspace{1.5mm}
           \mathrm{e}^{(-\lambda_{DC\_AC1} t)}  \times 
           \mathrm{e}^{(-\lambda_{DC\_AC2} t)}  \times 
           \mathrm{e}^{(-\lambda_{SA1} t)}  \times 
           \mathrm{e}^{(-\lambda_{SA2} t)})	\times \\ &\hspace{1.5mm}\mathrm{MTTR\_LoadA} \times \mathrm{CN\_LoadA} + \dots)          
\end{aligned}} {\mathrm{CN\_LoadA} +\mathrm{CN\_LoadB} + \mathrm{CN\_LoadC} + \mathrm{CN\_LoadD}}$$}}	          
\end{flushleft}

To further facilitate the exploitation of our proposed approach for power grid reliability engineers, we defined a Standard Meta Language (SML) functions~\cite{Etree_tp} that can numerically evaluate the above-\textit{verified} expressions of $\mathcal{FOR}_{PV}$, $\mathcal{FOR}_{STEAM}$,  and $\mathcal{SAIDI}$. Subsequently, we compared our results with MATLAB CCD algorithm based on Monte-Carlo Simulation~(MCS) and also with other existing subsystem-level reliability analysis techniques, such as HiP-HOPS and FMR, to ensure the accuracy of our computations, which is presented in the next~section.

\subsection{Experimental Results and Discussion}
Considering the failure rates of the power plant components $\lambda_{\mathrm{BO}}$, $\lambda_{\mathrm{TA}}$, $\lambda_{\mathrm{LF}}$, $\lambda_{\mathrm{DC\_DC}}$, $\lambda_{\mathrm{DC\_AC}}$ and $\lambda_{\mathrm{SA}}$ are 0.91, 0.84, 0.96, 0.67, 0.22, and 0.56 per year~\cite{li2013reliability}, respectively. Also, assuming that $\mathrm{MTTR}_{\mathrm{Load}_A}$, $\mathrm{MTTR}_{\mathrm{Load}_B}$, $\mathrm{MTTR}_{\mathrm{Load}_{C}}$, and $\mathrm{MTTR}_{\mathrm{Load}_{D}}$ are 12, 20, 15, and 10~hours/interruption~\cite{anders2011innovations} and  $\mathrm{CN}_{\mathrm{Load}_A}$, $\mathrm{CN}_{\mathrm{Load}_B}$, $\mathrm{CN}_{\mathrm{Load}_{C}}$, and $\mathrm{CN}_{\mathrm{Load}_{D}}$ are 500, 1800, 900, and 2500 customers, respectively. The reliability  study is undertaken for 1 year, i.e., \textit{t} = 8760 hours. Based on the given data, we can evaluate $\mathcal{FOR}$  and $\mathcal{SAIDI}$ for the electrical power network (Fig. \ref{Fig: Case study}) using following techniques:

\begin{enumerate}

\item Our proposed SML functions to evaluate the \textit{verified} expressions of $\mathcal{FOR}_{PV}$, $\mathcal{FOR}_{STEAM}$, and $\mathcal{SAIDI}$ in HOL4 (Theorems 23-25), as shown in Fig. \ref{Fig: HOL4 Results}.
\begin{figure} [!h]
	\begin{center}
		\centering
		\includegraphics[width = 0.8 \columnwidth]{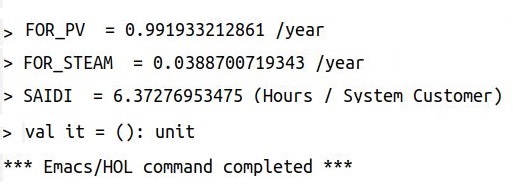} 
	\end{center}
	\caption{SML Functions: $\mathcal{FOR}$ and $\mathcal{SAIDI}$ Results}
	\label{Fig: HOL4 Results}
\end{figure}

\item MATLAB MCS-based toolbox that uses a random-based algorithm to obtain $\mathcal{FOR}$ and $\mathcal{SAIDI}$ for the electrical grid. The steps followed in this technique are~as follows~\cite{pradhan2020implementation}:

\begin{itemize}
\item Read the values of failure rate $\lambda$ in \textit{f/hours} and repair
time \textit{r} in hours for each component
\item Generate a random number \textit{U}
\item Calculate the predicted next Time to Fail (\textit{TTF}) and Time to Repair (\textit{TTR}) from the equations
\begin{equation}
    TTF = \frac{-\ln{U}}{\lambda} \hspace{4mm} TTR = \frac{-\ln{U}}{r}     
\end{equation}
\item Repeat the above iterative process till the number~of~iterations exceeds 1e5
\end{itemize}

Based on the above-mentioned MCS steps, we obtain different results of $\mathcal{FOR}$ and $\mathcal{SAIDI}$  every run of the algorithm depending on the generated random number with a tolerance  error between 4-9\%. So, we present in Table \ref{Table: CCD matlab results} the best-estimated results of $\mathcal{FOR}$ and $\mathcal{SAIDI}$ in MATLAB based on the MCS approach with the least errors. Subsequently, we take the mean average of all the obtained  $\mathcal{FOR}$ and $\mathcal{SAIDI}$ results for the power~grid. 

\begin{table} [!h]
 \caption{\protect  MATLAB MCS: $\mathcal{FOR}$ and $\mathcal{SAIDI}$ Results}
 \label{Table: CCD matlab results}
\begin{center}
 \begin{tabular}{|c|l|l|l|} 
 \hline \thead{Run} & \thead{$\mathcal{FOR}_{PV}$} & \thead{$\mathcal{FOR}_{STEAM}$} & \thead{$\mathcal{SAIDI}$} \\ [0.5ex] 
\hline \hline
 1 &  88.55e-2 &  36.18e-3  & 5.8023  \\  [1ex] 
\hline 
 2 & 107.19e-2  & 40.03e-3 & 6.5045  \\  [1ex] 
\hline 
 3 & 93.52e-2 & 36.35e-3   &  6.0222 \\  [1ex] 
\hline 
 5 & 110.17e-2  & 43.03e-3  & 7.0495  \\  [1ex] 
 \hline 
 4 & 95.24e-2  & 38.66e-3  & 6.3960 \\  [1ex] 
\hline 
Average &   98.93e-2 & 38.85e-3  & 6.3549 \\  [1ex] 
\hline 
\end{tabular}
\end{center}
\end{table} 

\item The Failure Mode Reasoning (FMR) approach, which identifies all the failure modes of safety-critical system inputs that can result in an undesired state at its output. The FMR process consists of four main stages  \cite{jahanian2019failure}: 
\vspace{-2mm}

\begin{enumerate}
    \item \textit{Composition}: Failure mode variables are defined and a set of logical implication statements  is generated that express local failure modes. 
    \item \textit{Substitution}: Local statements will be combined to create a single global implication statement between the critical-system inputs and outputs.
    \item \textit{Simplification}:  The complex formula is simplified, where we  trim off any redundant statements.
    \item \textit{Calculation}: The probability of failure is evaluated using the component failure rates.
\end{enumerate}

Based on the above-mentioned FMR procedures, we can express the component-level failure analysis of the PV power plant (Fig.~\ref{FT Model of a typical PV power plant}) as: 
\begin{equation}
\label{Equation 28} 
    (\hat{o} = \Dot{f}) \Rightarrow (\hat{x_1} = \Dot{f} \lor \hat{x_2} = \Dot{f})
\end{equation}

The above equation means that if the output $o$ is \textit{False} by
fault then either one of its inputs to the OR gate, i.e., $x_1$ or $x_2$, must be
\textit{False} by fault. We now need to determine what can cause
$\hat{x_1} = \Dot{f}$ and $\hat{x_2} = \Dot{f}$. Similar to Eq. 6, we can write:
\begin{equation}
\label{Equation 22} 
    (\hat{x_1} = \Dot{f}) \Rightarrow (\hat{x_3} = \Dot{f} \hspace{1mm} \lor \hat{x_4} = \Dot{f} \hspace{1mm} \lor \hat{x_5} = \Dot{f} \hspace{1mm} \lor \hat{x_6} = \Dot{f})
\end{equation}
\begin{equation}
\label{Equation 23} 
    (\hat{x_2} = \Dot{f}) \Rightarrow (\hat{x_7} = \Dot{f} \hspace{1mm} \lor \hat{x_8} = \Dot{f} \hspace{1mm} \lor \hat{x_9} = \Dot{f} \hspace{1mm} \lor \hat{x_{10}} = \Dot{f})
\end{equation}

where $x_3$, $x_4$, $x_5$, $x_6$, $x_7$, $x_8$, $x_9$, $x_{10}$ are $LF_1$, $DC\_DC_1$, $DC\_AC_1$, $SA_1$, $LF_2$, $DC\_DC_2$, $DC\_AC_2$, $SA_2$, respectively.    Similarly, we can express the component-level failure analysis of the steam~power~plant (Fig. \ref{Fig: FT Model of a typical steam power plant}) as: 
\begin{equation}
\label{Equation 24} 
    (\hat{o} = \Dot{f}) \Rightarrow (\hat{x_{11}} = \Dot{f} \hspace{0.5mm} \wedge  \hspace{0.5mm} \hat{x_{12}} = \Dot{f} \hspace{0.5mm} \wedge  \hspace{0.5mm} \hat{x_{13}} = \Dot{f})
\end{equation}
\begin{equation}
\label{Equation 25} 
    (\hat{x_{11}} = \Dot{f}) \Rightarrow (\hat{x_{14}} = \Dot{f} \hspace{0.5mm} \wedge  \hspace{0.5mm}  \hat{x_{15}} = \Dot{f})
\end{equation}
\begin{equation}
\label{Equation 26} 
    (\hat{x_{12}} = \Dot{f}) \Rightarrow (\hat{x_{16}} = \Dot{f} \hspace{0.5mm} \wedge  \hspace{0.5mm}  \hat{x_{17}} = \Dot{f})
\end{equation}
\begin{equation}
\label{Equation 27} 
    (\hat{x_{13}} = \Dot{f}) \Rightarrow (\hat{x_{18}} = \Dot{f} \hspace{0.5mm} \wedge  \hspace{0.5mm}  \hat{x_{19}} = \Dot{f})
\end{equation}

where $x_{14}$, $x_{15}$, $x_{16}$, $x_{17}$, $x_{18}$, $x_{19}$, are $BO_1$, $TA_1$, $BO_2$, $TA_2$, $BO_3$, $TA_3$, respectively. Table \ref{Table: CCD FMR results} shows the results of $\mathcal{FOR}_{PV}$,  $\mathcal{FOR}_{STEAM}$, and $\mathcal{SAIDI}$  based on FMR analysis using the assumed failure rates of the power plant components. 

\begin{table} [!h]
 \caption{\protect  FMR: $\mathcal{FOR}$ and $\mathcal{SAIDI}$ Results}
 \label{Table: CCD FMR results}
\begin{center}
 \begin{tabular}{|c|c|c|} 
 \hline \thead{$\mathcal{FOR}_{PV}$} & \thead{$\mathcal{FOR}_{STEAM}$} & \thead{$\mathcal{SAIDI}$} \\ [0.5ex] 
\hline\hline
99.19e-2 &  38.87e-3 & 6.3728 \\ [0.51ex]
\hline 
\end{tabular}
\end{center}
\end{table} 

According to Jahanian et al.~\cite{jahanian2020failure}, the soundness of the obtained FMR equations (Eq. \ref{Equation 28} to Eq. \ref{Equation 27}) needs to be proven mathematically.\\

\begin{figure} [!b]
	\begin{center}
		\centering
		\includegraphics[width = 0.6 \columnwidth]{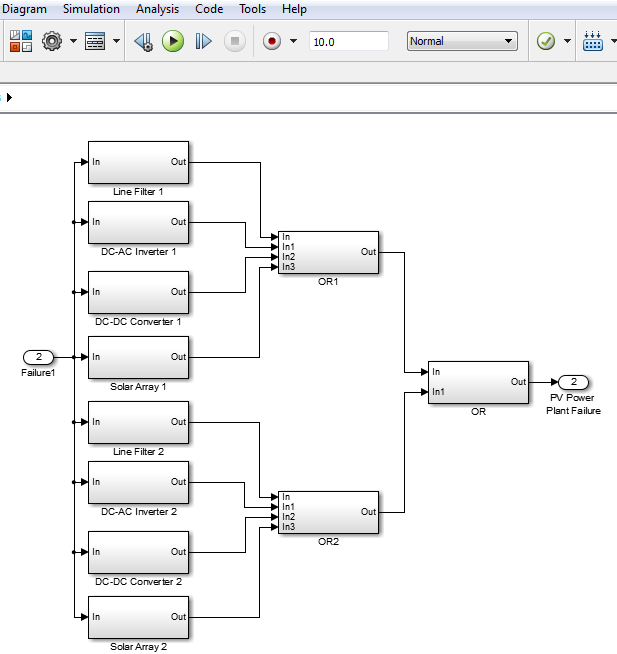} 
	\end{center}
	\caption{HiP-HOPS: PV Plant FMECA Analysis}
	\label{Fig: HIP-HOP2 Results}
\end{figure}

\item The HiP-HOPS software for failure analysis, which can perform FMECA analysis by given architectural blocks that hierarchically describe a safety-critical system at the subsystem level. Fig. \ref{Fig: HIP-HOP2 Results} and Fig. \ref{Fig: HIP-HOP1 Results} depict the  FMECA analysis of the PV and steam power plants using the HiP-HOPS software, respectively. The probabilistic results of $\mathcal{FOR}_{PV}$,  $\mathcal{FOR}_{STEAM}$, and $\mathcal{SAIDI}$  based on HiP-HOPS analysis are equivalent to the FMR analysis results presented~in~Table \ref{Table: CCD FMR results}.

\begin{figure} [!t]
	\begin{center}
		\centering
		\includegraphics[width = 0.7 \columnwidth]{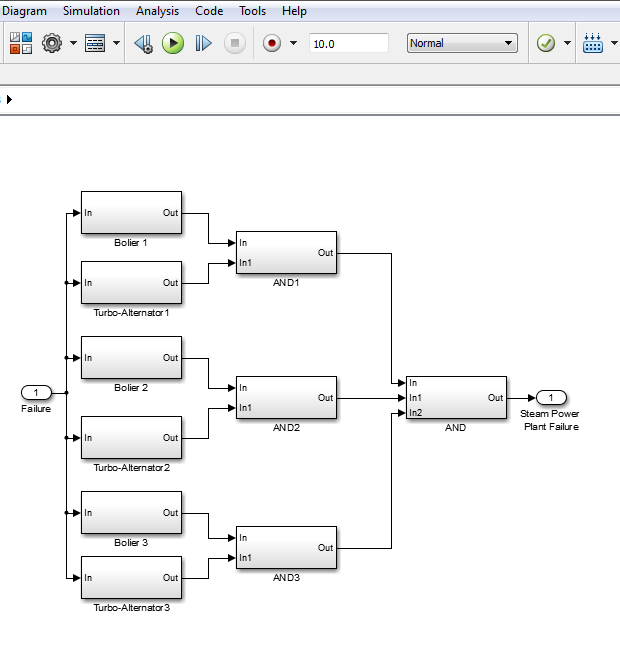} 
	\end{center}
	\caption{HiP-HOPS: Steam Plant FMECA Analysis}
	\label{Fig: HIP-HOP1 Results}
\end{figure}

\end{enumerate}

It can be observed that $\mathcal{SAIDI}$ result obtained from our formal HOL4 analysis are approximately equivalent to the corresponding ones calculated using FMR and HiP-HOPS approaches. On the other hand, MATLAB MCS-based uses a random-based algorithm, which estimates different results of $\mathcal{FOR}$ and $\mathcal{SAIDI}$ every generation of a random number with errors between 4-9\%. This clearly demonstrates that our analysis is not only providing the correct result but also with a \textit{formally proven} reliability expressions (Theorems~23-25) compared to simulation~tools, i.e.,~the soundness of subsystem-level reliability analysis. By performing the formal CCD step-analysis of a real-world 39-bus electrical power network, we demonstrated the practical effectiveness of the proposed CCD formalization in HOL4, which will help design engineers to meet the desired quality requirements. Also, our proposed formal approach can be used to analyze larger scale CCD models of other complex electrical power system applications, such as Smartgrids~\cite{fang2011smart}.


\section{Conclusions}
\label{Conclusions}
In this work, we developed a formal approach for Cause-Consequence Diagrams (CCD), which enables safety engineers to perform $\mathcal{N}$-level CCD analysis of safety-critical systems within the sound environment of the HOL4 theorem~prover. Our proposed approach provides new CCD mathematical formulations, which their correctness was verified in the HOL4 theorem prover. These formulations are capable of performing CCD analysis of \textit{multi-state} system components and based on any given probabilistic distribution and failure rates. These features are not available in any other existing approaches for subsystem-level reliability analysis. The proposed formalization is limited to perform CCD-based reliability analysis at the subsystem level that integrates static dependability analysis. However, this formalization is \textit{generic} and can be extended to perform dynamic failure analysis of dynamic subsystems where no dependencies exist in different subsystems. We demonstrated the practical effectiveness of the proposed CCD formalization by performing the formal CCD  step-analysis of a standard \textit{IEEE 39-bus} electrical power network system and also formally verified the power plants Force Outage Rate~($\mathcal{FOR}$) and the System Average Interruption Duration Index~($\mathcal{SAIDI}$). Eventually, we compared the $\mathcal{FOR}$ and $\mathcal{SAIDI}$ results obtained from our formal CCD-based reliability analysis with the corresponding ones using MATLAB based on Monte-Carlo~Simulation (MCS), the HiP-HOPS software tool, and the Failure Mode Reasoning (FMR) approach. As future work, we plan to integrate Reliability Block Diagrams (RBDs) \cite{ahmed2016formalization} as reliability functions in the CCD analysis, which will enable us to analyze hierarchical systems with different component success configurations, based on our CCD formalization in~the~HOL4 theorem prover. 

\bibliographystyle{IEEEtran}
\bibliography{Technical_Report}
\end{document}